\documentclass[reprint,amsmath,amssymb,aps,pre,showpacs,eqsecnum]{revtex4-1}

\usepackage{graphicx}      
\usepackage{dcolumn}       

    \usepackage[usenames]{color}

\newcommand{\zs}{{(\text{ZS})}}
\newcommand{\bbm}{\mathbf}

\newcommand{\cm}{\mathcal}
\newcommand{\beq}{\begin{equation}}
\newcommand{\eeq}{\end{equation}}
\newcommand{\beqn}{\begin{eqnarray}}
\newcommand{\eeqn}{\end{eqnarray}}
\def\bal#1\eal{\begin{align}#1\end{align}}

\newcommand\beqa{\begin{eqnarray}}
\newcommand\eeqa{\end{eqnarray}}
\newcommand{\nn}{\nonumber\\}
\newcommand{\dd}{\text{d}}
\newcommand{\hs}{\text{HS}}
\newcommand{\al}{\alpha}

\newcommand{\ed}{\end{document}}

\begin{document}
\title{Equation of state of sticky-hard-sphere fluids in the chemical-potential route}

\author{Ren\'e D. Rohrmann}
\email{rohr@icate-conicet.gob.ar}
\homepage{http://icate-conicet.gob.ar/rohrmann}
\affiliation{Instituto de Ciencias Astron\'omicas, de la Tierra y del
Espacio (ICATE-CONICET), Avenida Espa\~na 1512 Sur, 5400 San Juan, Argentina}

\author{Andr\'es Santos}
\email{andres@unex.es}
\homepage{http://www.unex.es/eweb/fisteor/andres/}

\affiliation{Departamento de F\'{\i}sica, Universidad de
Extremadura, Badajoz, E-06071, Spain}

\date{\today}
\begin{abstract}
The coupling-parameter method, whereby an extra particle is progressively coupled to the rest of the particles, is applied to the sticky-hard-sphere fluid to obtain
its equation of state in the so-called chemical-potential route ($\mu$ route).
{As a consistency test, the results} for one-dimensional sticky particles are shown to be exact. Results corresponding
to the three-dimensional case (Baxter's model)
are derived within the Percus--Yevick approximation  by using different prescriptions
for the dependence of the interaction potential of the extra particle on the coupling parameter.
The critical point and the coexistence curve of the gas-liquid phase transition
are obtained in the $\mu$ route and compared with predictions from other
thermodynamics routes and from computer simulations.
The results show that the $\mu$ route yields a general
better description than the virial, energy, compressibility, {and zero-separation} routes.

\end{abstract}



\pacs{
05.70.Ce, 	
61.20.Gy,   
61.20.Ne, 	
65.20.Jk 	
}

\maketitle
\section{ Introduction} \label{s.intr}

The chemical potential of a fluid can be evaluated as the change in the Helmholtz free energy when a new particle is added to the system through a coupling parameter \cite{O33,K35,H56,R80}.
The coupling parameter determines the strength of the interaction
of the added particle to the rest of the system and  usually varies between
zero (no interaction) and unity (full interaction).
This method provides the equation of state (EOS) of the fluid in the so-called chemical-potential route (or $\mu$ route).
This can be considered as the fourth route in addition to the better
known routes based on the pressure (or virial), energy, and compressibility
equations \cite{HM06}. It must be noted that all these ways to obtain the EOS are formally
equivalent.

In practice, the various thermodynamic routes have been mostly developed,
under the assumption of additive pair interactions, using the so-called radial
distribution function (RDF) $g(r)$. Within this class of interactions,
the evaluation of thermodynamic properties of a classical
fluid reduces to finding the corresponding RDF. Since all well-known
theoretical methods to obtain $g(r)$ give approximate solutions,
with the exception of a few, simple fluid models (for example,
one-dimensional systems whose particles interact only with their nearest
neighbors \cite{SZK53}), the EOS obtained from different routes differ in general
from one another.

The $\mu$ route has been largely unexplored, except in the scaled-particle
theory \cite{RFL59,LHP65,MR75,HC04b,SDC06}. Recently, one of us
used this method to obtain a hitherto unknown EOS for the hard-sphere (HS)
model in the Percus--Yevick (PY) approximation \cite{S12b}. This method was then extended to multicomponent fluids
for arbitrary dimensionality, interaction potential, and coupling protocol
\cite{SR13}. Its application to HS mixtures
allowed us to provide a new  EOS of this classical model in the PY approximation \cite{SR13} {and to derive the associated fourth virial coefficient in the hypernetted-chain approximation \cite{BS14}}.
Evidently, the $\mu$ route represents a helpful tool for the construction
of new EOS and the analysis of thermodynamic properties of fluids.
It is therefore of great interest to consider its application to
non-HS models.

In this paper we use the $\mu$ route to evaluate the EOS of the three-dimensional
sticky-hard-sphere (SHS) model introduced by Baxter \cite{B68}.
In this fluid, impenetrable particles of diameter $\sigma$ interact
through a square-well potential of infinite depth and vanishing width.
The study of this pair potential model has two advantages. First, it
admits an exact analytical solution to the Ornstein--Zernike equation with
the PY closure \cite{B68}, its thermodynamic properties being
 described in terms of two simple parameters, the packing fraction $\eta$
and a stickiness parameter $\alpha$ \cite{B74}.
Second, the SHS model has proved to provide an excellent starting point for the study of
colloidal systems with short-rang attraction \cite{CRST94,VD95,RZZ96,PFNR03,BRP07}, interactions between protein molecules in solution \cite{PPD98},
and other interesting applications \cite{NF00,MFGS13}.

We will exploit the known exact solution of the PY integral equation for both single and multicomponent SHS fluids \cite{B68,PS75} to obtain the EOS through the $\mu$ route and compare the outcome with the three standard routes (virial, energy, and compressibility), {with the less known zero-separation (ZS) route \cite{BT76,BT79},} and with Monte Carlo (MC) simulations \cite{MF04}.
As we will see, the $\mu$ route EOS in the PY approximation changes with the choice of the prescription followed to switch on the extra particle to the rest of the system. Despite this, the spread is typically much smaller than the one existing among the different routes. Interestingly enough, the $\mu$ route generally provides the best results, including the gas-liquid transition properties of the fluid.

The paper is organized as follows. In Sec.\ \ref{S2} we give the mathematical formulation of the $\mu$ route
for SHS fluids.
In Secs.\ \ref{S.shr} and \ref{S.shs} we use the known exact and PY solutions of the SHS  system in one and three dimensions, respectively to derive
the $\mu$ route EOS of
those systems. Section \ref{S.concl} is reserved for discussions of the results. The relevant calculations are presented in a series of appendixes.

\section{Chemical-potential route}\label{S2}

We consider a $d$-dimensional system of volumen $V$ containing $N=\rho V$ spherical particles
of diameter $\sigma$ with surface adhesion. The SHS interaction potential
$\phi(r)$ between two particles with centers separated  a distance $r$
is defined by
\beq \label{phi22}
e^{-\beta \phi(r)}=\Theta(r-\sigma) +\alpha \sigma \delta(r-\sigma),
\eeq
where $\beta=1/k_BT$, $k_B$ and $T$ being  the Boltzmann
constant and the absolute temperature, respectively. The dimensionless parameter $\alpha$ measures the strength of
surface adhesion (stickiness). The equality in (\ref{phi22}) must be interpreted as
the limit of an increasingly  deeper  ($\epsilon\to\infty$) and narrower ($\Delta\to 0$) square-well potential of depth $\epsilon$ and (relative) width $\Delta$  with
\beq
\alpha= e^{\beta\epsilon}\Delta
\label{alpha}
\eeq
kept constant.
The stickiness parameter $\alpha$ is related to the Baxter temperature $\tau$ \cite{B68} by $\tau=1/12\alpha$.
The pure HS model is recovered from Eq.\ (\ref{phi22}) in the
limit $\alpha\rightarrow 0$.

We now include into the system an additional particle (the solute). Its
interaction with any other particle in the fluid (the solvent) is given by
\beq \label{phi12}
e^{-\beta \phi_\xi(r)}=\Theta(r-\xi\sigma) + \alpha_\xi \xi\sigma\delta(r-\xi\sigma),
\eeq
where $\xi$ plays the role of a \emph{coupling} parameter
and  $\alpha_\xi$ is a continuous function of $\xi$ encoding the strength  of the solute-solvent attractive force. It  runs
from $\alpha_\xi=0$ at $\xi=0$ to $\alpha_\xi=\alpha $ at $\xi=1$.

For large $N$, the excess chemical potential
$\mu^{\rm ex}=\mu-\mu^{\rm id}$,  $\mu^{\rm id}$ being the contribution
of the corresponding ideal fluid, can be written as follows
\cite{K35,H56,R80,S12b,SR13}
\beq \label{mu0}
-\beta \mu^{\rm ex}= \ln \frac{Q_{N+1}^{(\xi=1)}}{Q_{N+1}^{(\xi=0)}},
\eeq
where
\beq \label{QN1a0}
Q_{N+1}^{(\xi)}=\frac{1}{V^{N+1}}\int \dd\bbm{r}^N\int\dd\bbm{r}_0\,
       e^{-\beta\Phi_{N+1}^{(\xi)}(\bbm{r}^{N+1})}
\eeq
is the configurational integral of $N$ solvent
particles plus one solute particle with a coupling parameter $\xi$.
Here, $\bbm{r}^{N+1}=\{\bbm{r}^N,{\bbm r}_0\}$, where $\bbm{r}^N$ refers to all
the translational coordinates of the $N$ solvent particles and ${\bbm r}_0$ refers to
the coordinates of the solute particle.
Furthermore,
\beq
\Phi_{N+1}^{(\xi)}(\bbm{r}^{N+1})
      =\frac{1}{2}\sum_{i\ne j}^N\phi(r_{ij})+\sum_{i=1}^N\phi_\xi(r_{0i})
\eeq
is the total potential energy, $r_{ij}$ being the distance between particles $i$ and $j$. Hence,
\beq \label{QN1a}
Q_{N+1}^{(\xi)}=\frac{1}{V^{N+1}} \int \dd\bbm{r}^N  e^{-\beta\Phi_{N}(\bbm{r}^N)}
     \int \dd\bbm{r}_0\prod_{i=1}^{N} e^{-\beta \phi_\xi(r_{0i})},
\eeq
where $\Phi_{N}(\bbm{r}^{N})$ denotes the solvent potential energy.

For convenience, we decompose the right-hand side of Eq.\ (\ref{mu0}) into
two separate contributions,
\beq \label{mu}
-\beta \mu^{\rm ex}= \ln \frac{Q_{N+1}^{(\frac{1}{2})}}{Q_{N+1}^{(0)}}
 +\int_{\frac{1}{2}}^1 \dd\xi\, \frac{\partial\ln Q_{N+1}^{(\xi)}}{\partial \xi}.
\eeq
Henceforth, for the sake of simplicity, we adopt $\sigma=1$. The two contributions in Eq.\ \eqref{mu} are worked out as follows.
First, with Eq.\ (\ref{phi12}), we write
\bal \label{ephi}
 \prod_{i=1}^{N} e^{-\beta \phi_\xi(r_{0i})}
  =&\prod_{i=1}^{N} \Theta(r_{0i}-\xi)\nn
  &+\alpha_\xi\xi\sum_{i=1}^{N} \delta(r_{0i}-\xi)
  \prod_{j\ne i} \Theta(r_{0j}-\xi)\nn
  &+(\alpha_\xi\xi)^2\sum_{i\ne j}^{N} \delta(r_{0i}-\xi)\delta(r_{0j}-\xi)\nn
  &\times\prod_{k\ne i,j} \Theta(r_{0k}-\xi) + \cm{O}(\alpha_\xi^3).
\eal
For $\xi<\frac{1}{2}$, the $N$ surfaces defined by $r_{0i}=\xi$ ($i=1,\ldots,N$) do not overlap, so that the condition $r_{0i}=\xi$ implies
$r_{0j}>\xi$ $\forall j\ne i$. As a consequence, the integrals in (\ref{QN1a}) of order two or higher in $\alpha_\xi$
vanish.
On the other hand, integration of $\prod_{i=1}^{N} \Theta(r_{0i}-\xi)$
over $\bbm{r}_0$ gives the free volume of the solute particle,
\beq
\int\dd\bbm{r}_0 \prod_{i=1}^{N} \Theta(r_{0i}-\xi)=V-N\Omega_\xi, \quad \xi<\frac{1}{2},
\eeq
where $\Omega_\xi=[\pi^{d/2}/{\Gamma (1+d/2)}]\xi^d$ is the volume of a $d$ sphere of radius $\xi$.
Furthermore,
\beq
\int  \dd\bbm{r}_0 \,\delta(r_{0i}-\xi) \prod_{j\ne i}
 \Theta(r_{0j}-\xi) = \Sigma_\xi, \quad \xi<\frac{1}{2},
\eeq
$\Sigma_\xi=\partial\Omega_\xi/\partial \xi=d\Omega_\xi/\xi$ being the surface of a $d$ sphere of radius $\xi$. Therefore,
\beq
\int \dd \bbm{r}_0 \prod_{i=1}^{N}  e^{-\beta \phi_\xi(r_{0i})}
  = V-N\Omega_\xi+N\alpha_\xi\xi \Sigma_\xi ,\quad \xi<\frac{1}{2}.
\eeq
With this result, Eq.\ (\ref{QN1a}) yields
\beq \label{QN1b}
Q_{N+1}^{(\xi)}= \left[ 1-\rho\Omega_\xi \left(1-d\alpha_\xi\right)\right] Q_N, \quad \xi<\frac{1}{2},
\eeq
where
\beq \label{QN}
Q_{N}= \frac{1}{V^N}\int \dd\bbm{r}^N  e^{-\beta\Phi_{N}(\bbm{r}^N)}
\eeq
is the configurational integral of the solvent.
As shown in Appendix \ref{appA}, the case $\xi=\frac{1}{2}$ is singular if $\alpha_{\frac{1}{2}}\neq 0$.
This difficulty can be overcome by the choice $\alpha_{\frac{1}{2}}= 0$. Therefore,
taking into account that $Q_{N+1}^{(0)}= Q_N$, and taking the limit $\xi\to \frac{1}{2}$ in Eq.\ \eqref{QN1b}, the first
term on the right-hand side of Eq.\ (\ref{mu}) becomes
\beq \label{mua}
\ln \frac{Q_{N+1}^{(\frac{1}{2})}}{Q_{N+1}^{(0)}}=
\ln\left( 1-\eta\right),
\eeq
where
\beq
\eta\equiv \rho\Omega_{\frac{1}{2}}
\label{eta}
\eeq
is the packing fraction.

For $\xi>\frac{1}{2}$ one must follow another strategy because integration over
$\bbm{r}_0$ in Eq.\ (\ref{QN1a}) depends upon the coordinates of all the solvent
particles. In this case, we consider
\bal \label{PN1b}
\frac{\partial e^{-\beta\Phi_{N+1}^{(\xi)}(\bbm{r}^{N+1})}}{\partial \xi}
 &=
e^{-\beta\Phi_{N}(\bbm{r}^{N})}
\frac{\partial \prod_{i=1}^{N}e^{-\beta \phi_\xi(r_{0i})}}
  {\partial \xi} \nn
&= e^{-\beta\Phi_{N}(\bbm{r}^{N})}  \sum_{i=1}^{N}
   \frac{\partial e^{-\beta \phi_\xi(r_{0i})}}{\partial \xi}
    \prod_{j\ne i}^{N} e^{-\beta \phi_\xi(r_{0j})} \nn
&= e^{-\beta\Phi_{N+1}^{(\xi)}(\bbm{r}^{N+1})} \sum_{i=1}^{N}
  e^{\beta \phi_\xi(r_{0i})}
\frac{\partial e^{-\beta \phi_\xi(r_{0i})}}{\partial \xi}.
\eal

The solute-solvent RDF  is  expressed as \cite{H56}
\beq \label{g2}
 g_\xi(r_{01})=\frac{V^{-(N-1)}}{Q_{N+1}^{(\xi)}} \int \dd\bbm{r}_2\cdots\int \dd\bbm{r}_N\, e^{-\beta\Phi_{N+1}^{(\xi)}(\bbm{r}^{N+1})}.
\eeq
It follows from Eqs.\ (\ref{QN1a}), (\ref{PN1b}), and (\ref{g2}) that
\bal \label{Qx2}
 \frac{\partial \ln Q_{N+1}^{(\xi)}}{\partial \xi}
 &=\frac {V^{-(N+1)}}{Q_{N+1}^{(\xi)}} \int \dd\bbm{r}^{N+1}
    \frac{\partial e^{-\beta\Phi_{N+1}^{(\xi)}(\bbm{r}^{N+1})}}{\partial \xi} \nn
 &= \frac1{V^2} \sum_{i=1}^{N} \int  \dd\bbm{r}_0\int \dd\bbm{r}_i\, y_\xi(r_{0i})
    \frac{\partial e^{-\beta \phi_\xi(r_{0i})}}{\partial \xi} \nn
 &=  \rho \int \dd \bbm{r} \, y_\xi(r)
    \frac{\partial e^{-\beta \phi_\xi(r)}}{\partial \xi}\nn
    &=d2^d\eta M_\xi(\eta,\alpha),
\eal
where
\beq \label{yxi}
 y_\xi(r)\equiv g_\xi(r)  e^{\beta \phi_\xi(r)}
\eeq
is the solute-solvent cavity function,
\beq
M_\xi(\eta,\alpha)\equiv\int_0^\infty \dd{r} \, r^{d-1}y_\xi(r)
    \frac{\partial e^{-\beta \phi_\xi(r)}}{\partial \xi},
    \label{Mxi}
\eeq
and in the last step of Eq.\ \eqref{Qx2} we have used spherical coordinates.

Finally, inserting Eqs.\ \eqref{mua} and \eqref{Qx2} into Eq.\ \eqref{mu}, we obtain
\beq \label{muex}
\beta \mu^{\rm ex}(\eta,\alpha)= -\ln\left( 1-\eta\right)
                    -d2^d\eta \int_{\frac{1}{2}}^1   \dd \xi \,M_\xi(\eta,\alpha).
\eeq
This gives the excess chemical potential  of $d$-dimensional SHS fluids as
obtained from the coupling parameter procedure.
To have an expression for $M_\xi$  more explicit than Eq.\ \eqref{Mxi}, we note that, according to Eq.\ \eqref{phi12},
\beq
\frac{\partial e^{-\beta \phi_\xi(r)}}{\partial \xi}=\frac{\partial \left(\alpha_\xi-1\right)\xi}{\partial\xi}\delta(r-\xi)-\alpha_\xi \xi\frac{\partial \delta(r-\xi)}{\partial r}.
\label{1}
\eeq
Thus,
\beq
M_\xi(\eta,\alpha)=
    \frac{\partial \left(\alpha_\xi-1\right)\xi}{\partial\xi}
   \xi^{d-1} y_\xi(\xi)+
    \alpha_\xi \xi\left. \frac{\partial \left[r^{d-1} y_\xi(r)\right]}{\partial r}
   \right|_{r=\xi}  .
   \label{fex}
\eeq

We may now derive the compressibility factor $Z\equiv \beta p/\rho$ ($p$ being the pressure) in the $\mu$ route.
The familiar thermodynamic relation
\beq
\left(\frac{\partial p}{\partial \rho}\right)_{T}=
\rho\left(\frac{\partial\mu}{\partial \rho}\right)_{T},
\eeq
can be expressed as
\beq \label{e.dZmu}
\frac{\partial [\eta (Z-1)]}{\partial \eta}=
\eta \frac{\partial(\beta\mu^{\rm ex})}{\partial \eta},
\eeq
{so that}
\beq
{Z(\eta,\alpha)=1+\beta\mu^{\text{ex}}(\eta,\alpha)-\int_0^1dt\,\beta\mu^{\text{ex}}(\eta t,\alpha).}
\label{Zmu}
\eeq
Thus, making use of  Eq.\ (\ref{muex}),  we obtain
\bal
\label{e.Zmu}
Z^{(\mu)}(\eta,\alpha) =&-\frac{\ln(1-\eta)}{\eta} -d2^d \eta \int_{\frac{1}{2}}^1   \dd \xi \Big[M_\xi(\eta,\alpha)\nn
   &  - \int_0^1\dd t \, tM_\xi(\eta t,\alpha)\Big].
\eal
This constitutes the EOS of $d$-dimensional SHS obtained from the
$\mu$ route (hence the superscript in $Z^{(\mu)}$). The better known virial, energy, and compressibility routes are worked out in Appendix \ref{A.vce}.

\section{Sticky hard rods: Exact results} \label{S.shr}
As a test of the correctness of Eq.\ \eqref{muex}, we prove in this section that it leads to the exact EOS for the one-dimensional system  ($d=1$).
In that case, Eq.\ \eqref{muex} reduces to
\beq \label{muexd1}
\beta \mu^{\rm ex}(\eta,\alpha) = -\ln(1 -\eta) -2\eta \int_{\frac{1}{2}}^1 \dd\xi M_\xi(\eta,\alpha)
\eeq
with
\beq
M_\xi(\eta,\alpha)=
   \frac{\partial \left(\alpha_\xi-1\right)\xi}{\partial\xi}
    y_\xi(\xi)
   + \alpha_\xi \xi y'_\xi(\xi)  .
   \label{fexd1}
\eeq
As shown in Appendix \ref{A.shr},
\beq  \label{yyd1}
 y_\xi(\xi)=\frac{\beta p/\eta}{1+\beta p\alpha_\xi\xi },\quad
 y'_\xi(\xi)=-\frac{(\beta p)^2/\eta}{1+\beta p\alpha_\xi\xi}.
\eeq
Thus, Eq.\ (\ref{fexd1}) may be written in the form
\beq
M_\xi(\eta,\alpha) = -\frac{\beta p}{\eta} + \frac{1}{\eta}\frac{\partial}{\partial_\xi }\ln(1+\beta p\alpha_\xi\xi).
\eeq
 Then, Eq.\ \eqref{muexd1} becomes
\beq \label{mushsd1}
\beta \mu^{\rm ex}(\eta,\alpha)= -\ln(1-\eta) +\beta p - 2\ln\left(1+\beta p\alpha\right).
\eeq
This result is exact and does not depend on the explicit form of
$\alpha_\xi$ in the interval $\frac{1}{2}<\xi<1$. Making use of Eq.\ \eqref{EOS_shr}, it is straightforward to check that Eq.\ \eqref{e.dZmu} is indeed satisfied.

\section{Sticky hard spheres: Percus--Yevick theory} \label{S.shs}

The excess chemical potential  for three-dimensional SHS fluids ($d=3$) is obtained from
Eqs.\  (\ref{muex}) and (\ref{fex})   as
\beq \label{muexd3}
\beta \mu^{\rm ex}(\eta,\alpha) = -\ln\left(1 -\eta \right) -24\eta
 \int_{\frac{1}{2}}^1 \dd\xi \,M_\xi(\eta,\alpha),
\eeq
\beq \label{fmud3}
M_\xi(\eta,\alpha) = \frac{\partial \left(\alpha_\xi-1\right)\xi}{\partial\xi}
   \xi^2 y_\xi(\xi) +\alpha_\xi \xi\left.\frac{\partial [r^2 y_\xi(r)]}
   {\partial r}\right|_{r=\xi}.
\eeq
The associated $\mu$ route compressibility factor is given by [see Eq.\ \eqref{e.Zmu}]
\bal
\label{e.Zmud3}
Z^{(\mu)}(\eta,\alpha) =&-\frac{\ln(1-\eta)}{\eta} -24 \eta \int_{\frac{1}{2}}^1   \dd \xi \Big[M_\xi(\eta,\alpha)\nn
   &   - \int_0^1 \dd t \,t M_\xi(\eta t,\alpha)\Big].
\eal

The evaluation of Eq.\  (\ref{fmud3}) requires the contact values of the solute-solvent cavity function $y_\xi(r)$ and its derivative $\partial_r y_\xi(r)$.
These may be obtained using the PY approximation for an SHS binary
mixture (see Appendix \ref{A.shs}). In particular, $y_\xi(\xi)$ and $y_\xi'(\xi)$ are given by Eqs.\  (\ref{yyd3x})
and (\ref{yyd3xc}), respectively.

\begin{figure}
\includegraphics[width=8cm]{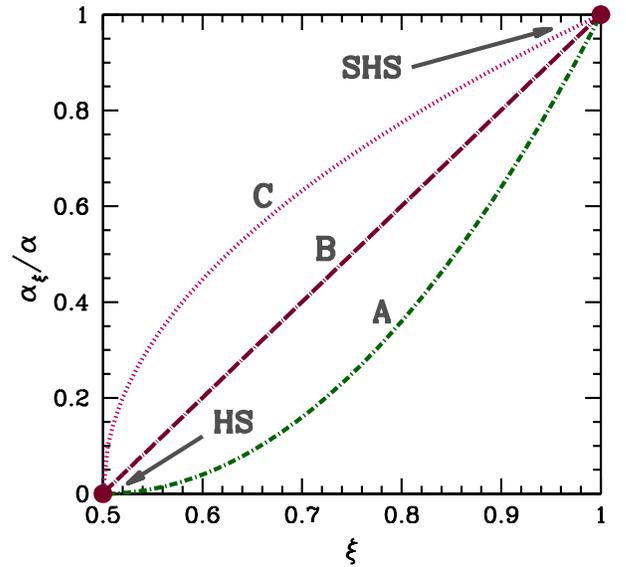}
  \caption{(Color online) The $\alpha_\xi$ stickiness parameter scaled by
$\alpha$, in the prescriptions A, B, and C given by Eq.\
(\ref{shs.ABC}).}
  \label{f.proto}
\end{figure}

For an explicit evaluation of Eqs.\ (\ref{muexd3})--(\ref{e.Zmud3}) we need to specify
the $\xi$-dependence of $\alpha_\xi$ (within the constraints $\alpha_{\frac{1}{2}}=0$ and $\alpha_1=\alpha$).
In this paper, we shall consider results based on three representative prescriptions:
\beq \label{shs.ABC}
\alpha_\xi=\begin{cases}
 (2\xi-1)^2    \alpha,& \text{(A)},\\
 (2\xi-1)      \alpha,& \text{(B)},\\
  \sqrt{2\xi-1}\alpha,& \text{(C)}.
\end{cases}
\eeq
These three protocols are depicted in Fig.\ \ref{f.proto}. In all of them,
the solute-solvent stickiness  monotonically grows from zero
to the solvent-solvent value as the solute diameter ($2\xi-1$) grows
from zero to the solvent diameter ($\sigma=1$). At a given solute
diameter, the strength of the solute-solvent attraction increases when going from A to C.

Since the PY integral equation is an \emph{approximate} theory, a common RDF is expected to yield different  EOSs depending on the route followed. In the case of the $\mu$ route, as will be seen below, an extra source of thermodynamic inconsistency arises: the EOS depends on the choice for the protocol $\alpha_\xi$.

\subsection{Virial expansion}\label{s.virial}

\begin{table*}
   \caption{Numerical values of the coefficients $b_4^\hs$ and $b_{4,i}$ [cf.\ Eq.\ \protect\eqref{b4i}].}\label{t.b4i}
\begin{ruledtabular}
\begin{tabular}{lccccccc}
 &$b_4^\hs$&$b_{4,1}$&$b_{4,2}$&$b_{4,3}$&$b_{4,4}$&$b_{4,5}$&$b_{4,6}$ \\  \hline
Exact&$18.36477$&$-165.283$&$880.416$&$-2623.10$&$3607.65$&$-1576.39$&$-194.468$\\
$v$&$16$&$-144$&$864$&$-2784$&$4032$&$-2304$&$0$\\
$e$&Undetermined&$-144$&$756$&$-2448$&$3888$&{$-2073.6$}&$0$\\
$c$&$19$&$-171$&$864$&$-2448$&$3456$&$-1728$&$0$\\
{ZS}&${5}$&${-45}$&${216}$&${-1296}$&${5184}$&${-5184}$&${0}$\\
$\mu_{\text{A}}$&$16.75$&$-150.75$&$853.795$&$-2711.64$&$4097.84$&$-2208.61$&$0$\\
$\mu_{\text{B}}$&$16.75$&$-150.75$&$860.384$&$-2737.15$&$4094.45$&$-2234.15$&$0$\\
$\mu_{\text{C}}$&$16.75$&$-150.75$&$866.194$&$-2759.45$&$4090.70$&$-2261.45$&$0$\\
     \end{tabular}
 \end{ruledtabular}
 \end{table*}

\begin{figure}
\includegraphics[width=8cm]{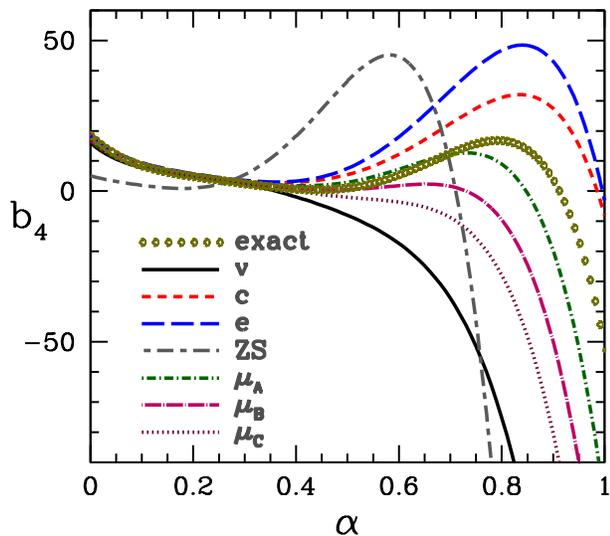}
\caption{(Color online) Comparison of the exact fourth virial coefficient  $b_4(\alpha)$  with the PY predictions in the virial ($v$),  energy ($e$), compressibility ($c$), {zero-separation (ZS),} and
chemical-potential ($\mu_{\text{A}}$, $\mu_{\text{B}}$, and $\mu_{\text{C}}$) routes.} \label{f.b4}
\end{figure}

\begin{table*}
   \caption{Numerical values of the coefficients $b_5^\hs$ and $b_{5,i}$ [cf.\ Eq.\ \protect\eqref{b5i}].}\label{t.b5i}
\begin{ruledtabular}
\begin{tabular}{lcccccccc}
 &$b_5^\hs$&$b_{5,1}$&$b_{5,2}$&$b_{5,3}$&$b_{5,4}$&$b_{5,5}$&$b_{5,6}$ &$b_{5,7}$\\  \hline
$v$&$22$&$-264$&$2700$&$-16\,920$&$63\,072$&$-134\,784$&$152\,064$&$-69\,120$\\
$e$&Undetermined&$-264$&$2160$&$-13\,104$&$51\,840$&$-120\,268.8$&$138\,240$&$-59\,245.7$\\
$c$&$31$&$-372$&$2916$&$-15\,048$&$50\,112$&$-100\,224$&$103\,680$&$-41\,472$\\
{ZS}&${-5.6}$&${67.2}$&${86.4}$&${- 3974.4}$&${29\,030.4}$&${- 124\,416}$&${248\,832}$&${- 165\,888}$\\
$\mu_{\text{A}}$&$23.8$&$-285.6$&$2680.78$&$-16\,322.8$&$61\,654.4$&{$-135\,696.4$}&{$152\,203.3$}&$-65\,563.1$\\
$\mu_{\text{B}}$&$23.8$&$-285.6$&$2715.20$&$-16\,592.7$&$62\,400.3$&{$-136\,366.0$}&{$153\,001.7$}&$-66\,620.8$\\
$\mu_{\text{C}}$&$23.8$&$-285.6$&$2745.36$&$-16\,831.5$&$63\,063.1$&{$-136\,973.3$}&{$153\,850.2$}&$-67\,765.8$\\
     \end{tabular}
 \end{ruledtabular}
 \end{table*}

\begin{figure}
\includegraphics[width=8cm]{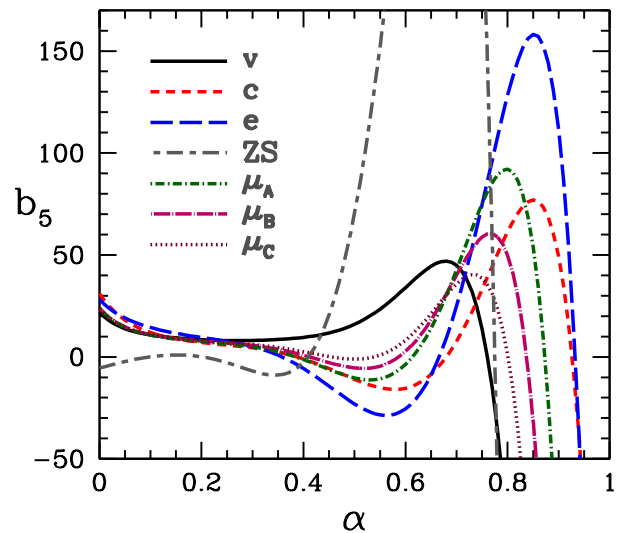}
\caption{(Color online) Fifth virial coefficient $b_5(\alpha)$ as predicted by the PY theory in the virial ($v$), energy ($e$), compressibility ($c$), {zero-separation (ZS),} and
chemical potential ($\mu_{\text{A}}$, $\mu_{\text{B}}$, and $\mu_{\text{C}}$)
routes.}
\label{f.b5}
\end{figure}

A standard method of examining different approximations in statistical mechanics
is to compare the successive terms in the virial expansion
of the compressibility factor. For SHS fluids,
\beq
Z(\eta,\alpha)=1+\sum_{j=2}^\infty b_j(\alpha) \eta^{j-1}.
\label{Zvir}
\eeq
The virial coefficients $b_j(\alpha)$  in the virial, energy, compressibility, and chemical-potential routes can be respectively evaluated from Eqs.\ (\ref{Zv}), (\ref{Zc}), (\ref{Zu}), and (\ref{e.Zmud3}), complemented by the PY results summarized in Appendix \ref{A.shs}. {The virial coefficients corresponding to an additional route, the so-called ZS route \cite{BT76,BT79}, can be derived within the PY approximation from Eq.\ \eqref{ZZS}.}

{All the routes in the PY approximation yield the exact second virial coefficient:}
\beq
b_2(\alpha)= 4-12\alpha.
\eeq
{As for the third virial coefficient, its exact expression,}
\beq
b_3(\alpha)=10-60\alpha+144\alpha^2-96\alpha^3,
\eeq
{is recovered from the virial, energy, compressibility, and chemical-potential routes, but not from the ZS route. The ZS result is}
\beq
{b_3^\zs(\al)=-\frac{4}{3}+8\alpha+96\alpha^2-192\alpha^3,}
\eeq
{which is especially wrong in the HS limit ($\alpha\to 0$).}

The exact fourth virial coefficient is a sixth-degree polynomial in $\alpha$ \cite{PG86}, i.e.,
\beq
b_4(\alpha)=b_4^\hs+\sum_{i=1}^6 b_{4,i} \alpha^i,
\label{b4i}
\eeq
where the numerical coefficients are given by the first row of Table \ref{t.b4i}.
The PY predictions for $b_4(\alpha)$ depend on the thermodynamic route. They have the structure of Eq.\ \eqref{b4i}, except that $b_{4,6}=0$.
The corresponding numerical coefficients are displayed in Table \ref{t.b4i}.
Note that the energy route  is unable to fix the HS EOS, so that the coefficients $b_j^{\hs}$ remain undetermined in that route.
All the coefficients $b_{4,i}$ derived from the $\mu$ route are rational numbers although a limited number of digits is shown in Table \ref{t.b4i}. Note that the three protocols of the $\mu$ route agree in the values of $b_4^\hs=\frac{67}{4}$ and $b_{4,1}=-\frac{603}{4}$. However, the coefficients $b_{4,2}$--$b_{4,5}$ depend on the choice of $\alpha_\xi$. For an arbitrary function $\alpha_\xi$, they are
\beq
b_{4,2}^{(\mu)}=837\left[1+\frac{12}{31}\int_{\frac{1}{2}}^1\dd\xi\,\left(\xi^2-\frac{1}{4}\right)\frac{\xi^2\alpha_\xi^2}{\alpha^2}\right],
\label{b42}
\eeq
\beq
b_{4,3}^{(\mu)}=-2646\left[1+\frac{24}{49}\int_{\frac{1}{2}}^1\dd\xi\,\left(\xi^2+\frac{\xi^2\alpha_\xi}{6\alpha}-\frac{3}{8}\right)\frac{\xi^2\alpha_\xi^2}{\alpha^2}\right],
\eeq
\beq
b_{4,4}^{(\mu)}=4104\left[1-\frac{3}{19}\int_{\frac{1}{2}}^1\dd\xi\,\left(1-\frac{4\xi^2\alpha_\xi}{3\alpha}\right)\frac{\xi^2\alpha_\xi^2}{\alpha^2}\right],
\eeq
\beq
b_{4,5}^{(\mu)}=-2160\left(1+\frac{2}{5}\int_{\frac{1}{2}}^1\dd\xi\,\frac{\xi^4\alpha_\xi^3}{\alpha^3}\right).
\label{b45}
\eeq

The three protocols in Eq.\ \eqref{shs.ABC} have the common form $\alpha_\xi=(2\xi-1)^q\alpha$ with $q=0.5$, $1$, and $2$ for C, B, and A, respectively. Taking $q>0$ as a free parameter and using Eqs.\ \eqref{b42}--\eqref{b45}, it is possible to find the \emph{optimal} value of $q$ that makes $b_4^{(\mu)}(\alpha)=b_4^{\text{exact}}(\alpha)$ for a given value $\alpha>0.282$. For instance, the optimal values are $q=0.199$, $1.208$, $2.076$, $3.702$, and $4.997$ for $\alpha=0.3$, $0.5$, $0.7$, $0.9$, and $1$, respectively. For $\alpha<0.282$ the mathematical solutions of $b_4^{(\mu)}(\alpha)=b_4^{\text{exact}}(\alpha)$ are $q<0$, but these are nonphysical values violating the condition $\alpha_{\frac{1}{2}}=0$.

Figure \ref{f.b4} compares the exact $b_4(\alpha)$ with various PY routes, where the Carnahan--Starling (CS) \cite{HM06} value $b_4^\hs=18$ has been taken in the case of the energy route.
{A very poor behavior of the ZS route is observed.} In what concerns the other four routes, small deviations occur among them for low and moderate stickiness ($\alpha\lesssim 0.35$), a good agreement with the exact values being found in that range. For $\alpha\gtrsim 0.35$, however, larger discrepancies occur,
with the energy and virial routes showing the most extreme
deviations with respect to the exact solution. The $\mu$ route predictions lie between the
virial and the compressibility ones, becoming closer to the exact values as a softer stickiness prescription is used (protocol A).

Although, to the best of our knowledge, the fifth virial coefficient is not exactly known, it is worthwhile comparing the different PY  predictions for it. They have the polynomial structure
\beq
b_5(\alpha)=b_5^\hs+\sum_{i=1}^7 b_{5,i} \alpha^i,
\label{b5i}
\eeq
the coefficients being presented in Table \ref{t.b5i}. Again, all the coefficients are rational numbers. The dependence of $b_5(\alpha)$ on the stickiness parameter is shown in Fig.\ \ref{f.b5} (with the CS choice $b_5^\hs=28$ for the energy route). As expected, the influence of the thermodynamic route on $b_5(\alpha)$ is stronger than in the case of $b_4(\alpha)$. The general shapes  of $b_5$ in the $\mu$  and compressibility routes are intermediate between those in the virial and energy routes.

\subsection{Weakly sticky limit}\label{s.alphalow}

\begin{table*}
   \caption{Expressions for $Z_\hs(\eta)$ and the coefficients ${Z}_1(\eta)$ and ${Z}_2(\eta)$ [cf.\ Eq.\ \protect\eqref{ci}].}\label{t.ci}
\begin{ruledtabular}
\begin{tabular}{lccc}
 &$Z_\hs(\eta)$&${Z}_{1}(\eta)$&${Z}_{2}(\eta)$\\  \hline
$v$&$\frac{1+2\eta+3\eta^2}{(1-\eta)^2}$&$
       -\frac{12\eta(1+2\eta)}{(1-\eta)^3}$&$\frac{36\eta^2(2+\eta)(2+3\eta)}{(1-\eta)^4}$\\
$e$&Undetermined&$-\frac{12\eta(1+2\eta)}{(1-\eta)^3}$&$\frac{36\eta^2(4+5\eta)}{(1-\eta)^4}$\\
$c$&$\frac{1+\eta+\eta^2}{(1-\eta)^3}$&$
       -\frac{3\eta(2+\eta)^2}{(1-\eta)^4}$&$\frac{36\eta^2(2+\eta)^2}{(1-\eta)^5}$\\
{ZS}&${-\frac{\ln\left[(1+2\eta)(1-\eta)^4\right]}{\eta}-1}$&${-3\frac{\ln\left[(1+2\eta)(1-\eta)^4\right]}{\eta}-\frac{6(1+5\eta)}{(1-\eta)(1+2\eta)}}$
&${-3\frac{\ln\left[(1+2\eta)^5(1-\eta)^{64}\right]}{\eta}-\frac{18(9+25\eta-17\eta^2-71\eta^3)}{(1-\eta)^2(1+2\eta)^2}}$\\
$\mu_{\text{A}}$&$-9\frac{\ln(1-\eta)}{\eta}-\frac{16-31\eta}{2(1-\eta)^2}$&$-27\frac{\ln(1-\eta)}{\eta}-\frac{3(18-37\eta+49\eta^2)}{2(1-\eta)^3}$&
{$-648\frac{\ln(1-\eta)}{\eta}-\frac{9(8064-28\,224\eta+33\,152\eta^2-20\,257\eta^3)}{112(1-\eta)^4}$}\\
$\mu_{\text{B}}$&$-9\frac{\ln(1-\eta)}{\eta}-\frac{16-31\eta}{2(1-\eta)^2}$&$-27\frac{\ln(1-\eta)}{\eta}-\frac{3(18-37\eta+49\eta^2)}{2(1-\eta)^3}$&
{$-\frac{96\,363}{140}\frac{\ln(1-\eta)}{\eta}-\frac{27(7138-24\,983\eta+29\,438\eta^2-17\,820\eta^3)}{280(1-\eta)^4}$}\\
$\mu_{\text{C}}$&$-9\frac{\ln(1-\eta)}{\eta}-\frac{16-31\eta}{2(1-\eta)^2}$&$-27\frac{\ln(1-\eta)}{\eta}-\frac{3(18-37\eta+49\eta^2)}{2(1-\eta)^3}$&
{$-\frac{28\,917}{40}\frac{\ln(1-\eta)}{\eta}-\frac{9(6426-22\,491\eta+26\,566\eta^2-15\,967\eta^3)}{80(1-\eta)^4}$}\\
     \end{tabular}
 \end{ruledtabular}
 \end{table*}

\begin{figure}
\includegraphics[width=8cm]{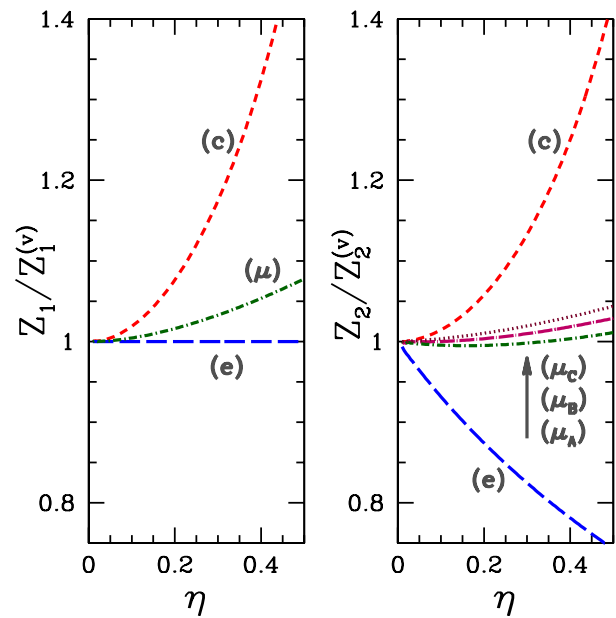}
\caption{(Color online) Coefficients $Z_1(\eta)$ and $Z_2(\eta)$  in
the $\alpha$ expansion of $Z(\eta,\alpha)$ [cf.\ Eq.\ \protect\eqref{ci}] from the PY equation in the
 energy ($e$), compressibility ($c$), and chemical-potential ($\mu_{\text{A}}$,
$\mu_{\text{B}}$, and $\mu_{\text{C}}$) routes, relative to the coefficients $Z_1^{(v)}(\eta)$ and $Z_2^{(v)}(\eta)$ obtained in the virial route.}
\label{f.dd}
\end{figure}

As a complement of the virial expansion \eqref{Zvir}, it is of interest to examine the leading terms in the  series expansion
\beq
Z(\eta,\alpha)=Z_\hs(\eta)+\sum_{i=1}^\infty {Z}_i(\eta)\alpha^i
\label{ci}
\eeq
of the compressibility factor in powers of the stickiness parameter.
Obviously, the zeroth-order coefficient in the $\alpha$ expansion is just the
compressibility factor of the pure HS system. Equation \eqref{ci} can be interpreted as a high-temperature expansion.

{}Making use of the  results of Appendix \ref{A.shs} in Eqs.\ (\ref{Zv}), (\ref{Zc}), (\ref{Zu}), {\eqref{ZZS},} and (\ref{e.Zmud3}), the first-order and second-order coefficients from the different routes can be derived. The results are displayed in Table \ref{t.ci}. Interestingly, one has $Z_1^{(v)}(\eta)=Z_1^{(e)}(\eta)$, thus generalizing the results $b_{4,1}^{(v)}=b_{4,1}^{(e)}$ and $b_{5,1}^{(v)}=b_{5,1}^{(e)}$ observed in Tables \ref{t.b4i} and \ref{t.b5i}. This reinforces that the natural extension of the energy route to HS fluids is the virial EOS \cite{S05,S06}.

The HS EOS $Z_\hs^{(\mu)}(\eta)$ was already derived in Ref.\ \cite{S12b}. We observe that $Z_1^{(\mu)}(\eta)$ is protocol-independent. This generalizes to any order in density the behavior observed for $b_{4,1}^{(\mu)}$ and $b_{5,1}^{(\mu)}$ in Tables \ref{t.b4i} and \ref{t.b5i}, respectively.
The expression of $Z_2^{(\mu)}(\eta)$ for arbitrary $\alpha_\xi$ is
\bal
Z_2^{(\mu)}(\eta)=&-540\frac{\ln(1-\eta)}{\eta}-\frac{18(30-105\eta+122\eta^2-77\eta^3)}{(1-\eta)^4}\nn
&-3888\int_{\frac{1}{2}}^1\dd\xi\,\xi^2\left(\xi-\frac{1}{2}\right)\left(\xi-\frac{1}{6}\right)\Bigg[\frac{\ln(1-\eta)}{\eta}
\nn
&
+\frac{1-\frac{7}{2}\eta+\frac{13}{3}\eta^2-\frac{13\xi-\frac{11}{6}}{6\xi-1}\eta^3}{(1-\eta)^4}
\Bigg]\frac{\alpha_\xi^2}{\alpha^2}.
\eal

The coefficients $Z_1(\eta)$ and $Z_2(\eta)$, relative to the virial-route predictions, obtained from various PY routes {(except the ZS one, in order to avoid distortion of the scales)} are shown in Fig.\ \ref{f.dd}. The discrepancies grow as density increases. In particular, the largest inconsistencies occur between the energy and compressibility routes. On the other hand, the $\mu$ route   deviates only slightly from the virial route, especially in the case of $Z_2$.

\subsection{Finite density and stickiness}

\begin{figure}
\includegraphics[width=8cm]{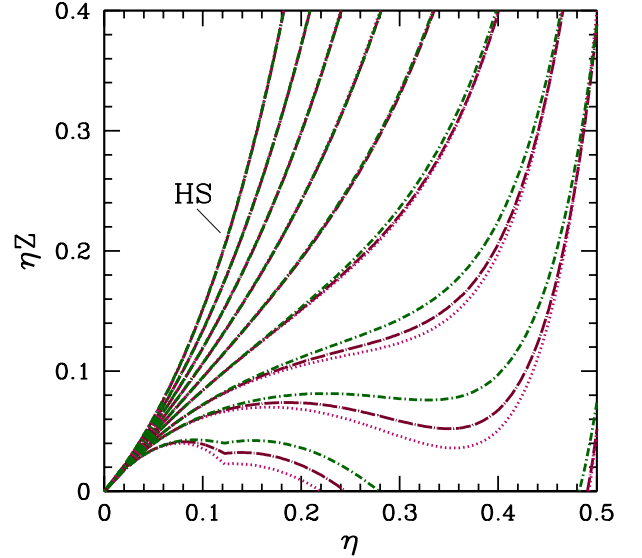}
\caption{(Color online) Reduced pressure $\eta Z^{(\mu)}$ of SHS fluids, as obtained
from the PY solution in the $\mu$ route according to the protocols A (-$\,\cdot\,$-$\,\cdot\,$-$\,\cdot$), B (---$\,\cdot\,$---$\,\cdot\,$---$\,\cdot$), and C ($\cdots\cdots$). The values of $\alpha$ are (from left to right) $\alpha=0$, $0.1$,
$0.2$, $0.3$, $0.4$, $0.5$, $\alpha_c^{(v)}\simeq 0.612\,418$, $\alpha_c^{(e)}\simeq 0.703\,209$, and
$\alpha_c^{(c)}\simeq 0.853\,553$.}
  \label{f.Zq}
\end{figure}

\begin{figure}
\includegraphics[width=8cm]{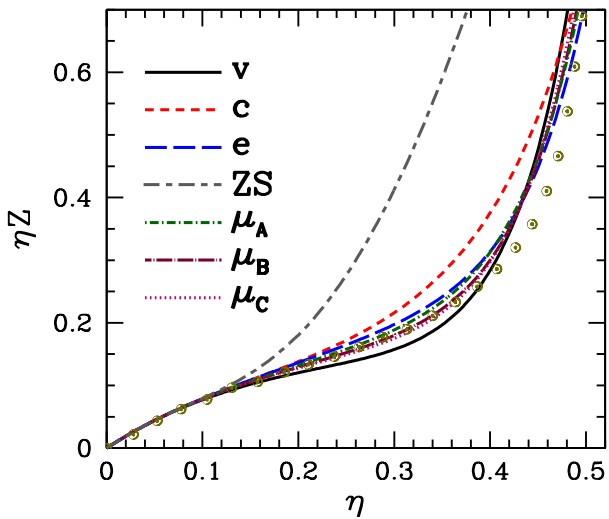}
\caption{(Color online) Reduced pressure $\eta Z$ as a function of the
packing fraction for SHS fluids at $\alpha=\frac{5}{9}\simeq 0.556$ ($\tau=\frac{3}{20}=0.15$).
The curves correspond to PY results from various routes as indicated on
the plot. Open circles represent MC calculations \cite{MF04}.}
  \label{f.Zmc}
\end{figure}

After having examined the low-density and low-stickiness (or high-temperature) regimes, we now consider the full non-perturbative regime.
The density dependence of the reduced pressure $\eta Z^{(\mu)}$
is plotted in Fig.\ \ref{f.Zq} for different values of $\alpha$  and for the three
protocols (\ref{shs.ABC}). Of course, in the limit of zero stickiness ($\alpha=0$), the choice of the protocol becomes irrelevant and one recovers the EOS $Z^{(\mu)}_{\rm HS}(\eta)$ (see Table \ref{t.ci}) corresponding to the PY theory in the $\mu$ route  \cite{S12b}.
As $\alpha$ increases, the pressure decreases with respect to the HS value and the influence of the protocol is practically negligible up to $\alpha\approx 0.5$. For higher stickiness, however, the values of $Z^{(\mu)}$ are increasingly sensitive to the protocol chosen. We observe that, as expected on physical grounds,  the stronger the relative stickiness $\alpha_\xi/\alpha$, the smaller the pressure.

The three higher values of $\alpha$ in Fig.\ \ref{f.Zq} correspond to the gas-liquid critical values $\alpha_c^{(v)}$, $\alpha_c^{(e)}$, and $\alpha_c^{(c)}$ predicted by the PY approximation in the virial, energy, and compressibility routes, respectively (see below).
In fact the kink in $Z^{(\mu)}$ at $\alpha=\alpha_c^{(c)}$ and $\eta=\eta_c^{(c)}\simeq 0.12132$ reflects the fact that $(\eta_c^{(c)},\alpha_c^{(c)})$ is the critical point for the existence of real solutions of the PY equation (see Appendix \ref{A.shs}).

In Fig.\ \ref{f.Zmc} we compare MC simulations at $\alpha=\frac{5}{9}$
\cite{MF04} with  PY predictions from {the different routes}.
Since, as discussed before, the energy route leaves the integration constant $Z_\hs(\eta)$ undetermined, henceforth the CS EOS \cite{HM06}
\beq
Z_\hs^{\text{CS}}(\eta)=\frac{1+\eta+\eta^2-\eta^3}{(1-\eta)^3}
\label{CS}
\eeq
will be taken to complete the determination of $Z^{(e)}(\eta,\alpha)$ via Eq.\ (\ref{Zu}), despite the fact that the choice $Z_\hs^{(e)}=Z_\hs^{(v)}$ would be more consistent \cite{S05,S06}. The
virial, compressibility, {and ZS} data have been obtained from Eqs.\ (\ref{Zv}),
(\ref{Zc}), {and \eqref{ZZS},} respectively. In all the cases, use has been made of the PY solution detailed in Appendix
\ref{A.shs}.

We observe that in the low-density range ($\eta\lesssim 0.15$) all PY routes and simulation data
agree very well. For higher densities, the {ZS pressure grows too rapidly} and the curves corresponding to the three different
protocols of the $\mu$ route remain rather close  in comparison with
those from the virial, energy, and compressibility routes, which show a larger
spread. In the range $0.2 \lesssim\eta \lesssim 0.4$, the $\mu$ route gives the best fits
to the simulation data. In the same region, $Z^{(e)}$ and $Z^{(c)}$ overestimate the simulation values, while $Z^{(v)}$ underestimates them. Up to $\eta\approx 0.4$, one has $Z^{(v)}<Z^{(\mu_{\text{C}})}<Z^{(\mu_{\text{B}})}<Z^{(\mu_{\text{A}})}<Z^{(e)}<Z^{(c)}$.
Finally, there is a rather strong disagreement of all the PY routes at high densities,
$0.4 \lesssim \eta \lesssim 0.5$, where the simulation data exhibit lower pressure values
than the theoretical ones. {Aside from the ZS curve,} the compressibility route shows the largest
deviations from MC results on the whole range of studied densities.

\subsection{Gas-liquid transition}

\begin{table}
   \caption{Comparison of the SHS gas-liquid critical values of $\alpha$, $\eta$, $\tau=1/12\alpha$,
and $\rho=6\eta/\pi$  from MC simulations
\cite{MF04} and PY solutions in the virial,
energy, compressibility, and  chemical-potential routes.}\label{t.critical}
\begin{ruledtabular}
\begin{tabular}{lccccccccc}
&MC& $v$&$e$&$c$ &{ZS}& $\mu_{\text{A}}$ & $\mu_{\text{B}}$ & $\mu_{\text{C}}$  \\  \hline
$\alpha_c$&$0.7355$ &$ 0.6124$ & $0.7032$ &$0.8536$ & ${0.7112}$ & {$0.6858$}& {$0.6605$}  &{$ 0.6412$} \\
$\tau_c$&$0.1133$ & $0.1361$ &  $0.1185$ &$0.0976$ &${0.1172}$ & {$0.1215$}& {$0.1262$}  & {$0.1300$} \\
$\eta_c$&$0.2660$ & $0.2524$ &  $0.3187$ &$0.1213$&${0.1039}$ & {$0.2761$}& {$0.2691$}  & {$0.2645$} \\
$\rho_c$&$0.5080$ & $0.4820$ &  $0.6086$ &$0.2317$&${0.1985}$ & {$0.5274$}& {$0.5140$}  & {$0.5051$} \\
     \end{tabular}
 \end{ruledtabular}
 \end{table}

\begin{figure}
\includegraphics[width=8cm]{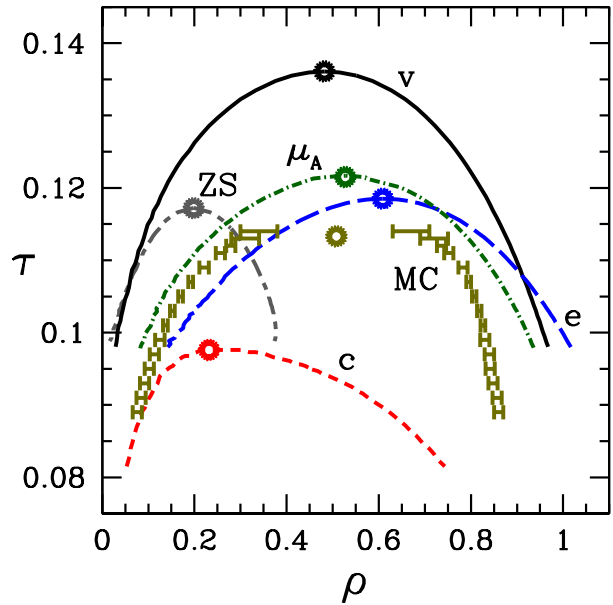}
\caption{(Color online) Phase diagram of the SHS fluid showing
the gas-liquid coexistence curves from PY solutions in the virial (---),  energy (-- -- --), compressibility (- - -), {ZS (-- - -- -)}, and chemical-potential (-$\,\cdot\,$-$\,\cdot\,$-$\,\cdot$) routes. Results in the $\mu$ route are based
on prescription A. MC simulation data taken from Ref.\ \protect\cite{MF04}
are shown with error bars. The critical points are indicated with circles.}
\label{f.crit}
\end{figure}

The two highest values of $\alpha$ in Fig.\  \ref{f.Zq} show a domain of mechanical instability,
$\partial(\eta Z)/\partial\eta<0$ (i.e., $\partial p/\partial\rho<0$),
which indicates a phase transition according to the $\mu$ route in the
various protocols analyzed. As is well known, the SHS fluid has a (metastable)
fluid-fluid transition which was early predicted in the compressibility
\cite{B68} and energy \cite{WHB71} routes of the PY approximation.
Critical values of the parameters  $\alpha$ and $\eta$  (or $\tau$ and $\rho$)
can be determined by the conditions $\partial p/\partial\rho=0$ and
$\partial^2 p/\partial\rho^2=0$.
Results concerning to various PY routes are summarized in Table
\ref{t.critical}, where they can be compared to the ones obtained by Miller
 and Frenkel \cite{MF04} using MC simulations.

As  is  known, the compressibility route produces a gross
underestimation of the critical density. {An even higher underestimation of $\eta_c$ is obtained from the ZS method.} The critical density is much better approximated by
the virial route (deviation of $5\%$) and, especially, the $\mu$ route (deviations of  $4\%$, $1.2\%$, and $0.6\%$ for protocols A, B, and C, respectively). On the other hand, the critical value of the
stickiness parameter evaluated from the virial and the compressibility routes
differ significantly from those predicted by numerical experiments.
For this parameter, {the ZS route (deviation of $3\%$)}, the energy route (deviation of $4\%$) and the $\mu$ route (deviations of  $7\%$, $10\%$, and $13\%$ for A, B, and C, respectively) give the best results.
{In view of the general poor performance of the ZS route, its good prediction of the critical stickiness can be viewed as accidental. In addition,} it must be remarked that the critical point obtained from $Z^{(e)}$ is quite sensitive to the choice of $Z_\hs$. If, instead of the CS EOS \eqref{CS},  the more consistent choice \cite{S05,S06} $Z_\hs=Z_\hs^{(v)}$ is used, then no SHS critical point is predicted by the energy route.
In conclusion, the parameters of the critical points obtained from the
$\mu$ route show the best global agreement with simulations.

We have also computed coexistence curves with the familiar
equal-area Maxwell construction that is applicable when $\partial p/\partial\rho<0$.
Figure \ref{f.crit} displays the  coexistence curve and the location of
the critical point derived by various PY routes and from
computer simulations \cite{MF04}. For simplicity, in the case of the $\mu$ route only
 results from protocol A are shown.
As may be seen in Fig.\ \ref{f.crit}, the curves obtained from the  virial,
compressibility, {and ZS} routes differ substantially from computer evaluations.
On the contrary, the agreement is reasonably good for the energy and
$\mu$  routes. As already seen from Table \ref{t.critical},  the critical Baxter temperature $\tau_c$ predicted by the  energy route is more accurate than the one obtained from the $\mu$ route. However, the general shape of the coexistence curve (both the gas and the liquid branches) is better described by the $\mu$ route.

\section{Conclusions}\label{S.concl}

In this paper we have studied the SHS fluid using the concept of a partially coupled
particle, whereby the interaction potential connecting this particle (the solute)
to all other particles  in the fluid (the solvent) is regulated by a charging
parameter $\xi$ which varies from $\xi=0$ (no interaction) to $\xi=1$
(full interaction). With this method, first introduced by Onsager \cite{O33} and
subsequently developed by Kirkwood \cite{K35} and other authors \cite{H56,RFL59}, we have derived the chemical potential  of  $d$-dimensional SHS fluids in terms of the contact values
of the solute-solvent cavity function $y_\xi(\xi)$ and its first derivative
$y'_\xi(\xi)$ [see Eqs.\ \eqref{fex} and \eqref{muex}].

The procedure requires $y_\xi(\xi)$ and $y'_\xi(\xi)$ in the range
$\frac12<\xi<1$, where the effective diameter ($2\xi-1$) of the coupled particle varies
between zero to the solute diameter $\sigma=1$. Thus, the explicit
evaluation of the EOS of the fluid in the $\mu$ route requires the structural  functions of the corresponding binary system in the infinite dilution limit.
While this may represent a practical disadvantage with respect to the other standard routes (which only require
the RDF of the one-component fluid), it nicely complements them. As is well known, the natural variables of the free energy $F$ are the temperature ($T$), the volume ($V$), and the number of particles ($N$). The internal energy, the pressure and the isothermal compressibility are directly related to $\partial F/\partial T$, $\partial F/\partial V$, and $\partial^2 F/\partial V^2$, respectively. {}Therefore, the chemical potential, being related to $\partial F/\partial N$, completes the picture.

The $\mu$ route also requires a prescription for the solute-solvent
interaction potential, i.e.,  the dependence of the solute-solvent
stickiness $\alpha_\xi$ on the coupling parameter $\xi$ must be specified. Here, in order to avoid the effects of trimer particle configurations, we have selected
prescriptions with $\alpha_\xi=0$ at $ \xi= \frac12$.
The resulting EOS, Eq.\ (\ref{e.dZmu}), is exact if the correct contact values $y_\xi(\xi)$ and $y'_\xi(\xi)$ are used,  regardless of
the explicit form of $\alpha_\xi$ in the range $\frac 12<\xi\le 1$.
This has been checked for the one-dimensional sticky fluid, in which case the exact
EOS is recovered from the $\mu$ route (see Sec.\ \ref{S.shr}).

We have also applied the $\mu$ route to three-dimensional SHS fluids in the
PY approximation. In this case, since the associated RDF is only approximate, the $\mu$ route EOS is influenced by the choice of the $\alpha_\xi$-protocol. On the other hand, this thermodynamical inconsistency becomes small in comparison with the spread of results obtained from the other  routes
(virial, energy,  compressibility, {and  ZS}).
When compared with available simulation data \cite{MF04}, the $\mu$ route EOS exhibits a general better agreement  than those evaluated from the other routes. The gas-liquid phase transition has also been analyzed. The $\mu$ route provides the best prediction for the critical density, being only improved by the {ZS and} energy routes
in the prediction of the critical stickiness parameter. To put this latter fact in perspective, it is important to remark that, as usually done \cite{WHB71}, the energy route has been complemented \emph{ad hoc} by the accurate CS EOS for the HS fluid. As for the coexistence curve, the best global agreement with simulation data is obtained from the $\mu$ route.
In addition, comparison of the fourth virial coefficient with exact results shows that the best performance corresponds to the $\mu$ route with a slower switching on of stickiness (protocol A).

To conclude, we expect that the results presented in this paper may contribute to place the $\mu$ route on the same footing as the other three conventional routes. This is especially important in the case of mixtures \cite{SR13}, where the chemical-potential concept fits in a more natural way. Regarding the PY approximation, it is interesting to note that, whereas in the case of the HS fluid \cite{S12b,SR13} the best behavior corresponds to the compressibility route (followed by the $\mu$ route), the inclusion of an attractive part in the interaction potential seems to favor the $\mu$ route as the most advantageous one.

\acknowledgments
{R.D.R. is grateful to the Consejo Nacional de Investigaciones
Cient\'ificas y T\'ecnicas (CONICET, Argentina) by financial support through Grants Nos.\
PIP 112-200801-01474 and PIP 114-201101-00208. The research of A.S. has been supported by the Spanish Government and by the Junta de Extremadura (Spain) through Grants Nos.\ FIS2010-12587 and  GRU10158, respectively, both partially
financed by Fondo Europeo de Desarrollo Regional (FEDER) funds.}

\appendix

\section{The limit $\xi\to \frac{1}{2}$}
\label{appA}

In contrast to the situation with $\xi<\frac{1}{2}$, a solute particle with $\xi=\frac{1}{2}$ allows spatial configurations where the solute and two solvent particles are simultaneously touching each other.
Such configurations  have a non-zero statistical weight
in the evaluation of Eq.\ (\ref{QN1a}) through the term of order $\alpha_\xi^2$
in Eq.\ (\ref{ephi}), unless $\alpha_{\frac{1}{2}}=0$.  Such a term is
\bal
Q_{N+1}^{(\frac{1}{2},2)}\equiv&
\frac{\alpha_{\frac{1}{2}}^2}{4V^{N+1}}\int\dd \bbm{r}^Ne^{-\beta\Phi_N(\bbm{r}^N)}\int\dd\bbm{r}_0\sum_{i\ne j}^{N} \delta\left(r_{0i}-\frac{1}{2}\right)\nn &
\times\delta\left(r_{0j}-\frac{1}{2}\right)\prod_{k\ne i,j} \Theta\left(r_{0k}-\frac{1}{2}\right)\nn
=&\alpha_{\frac{1}{2}}^2\frac{N(N-1)}{4V^{N+1}}\int\dd \bbm{r}^N e^{-\beta\Phi_N(\bbm{r}^N)}\nn
&\times \int\dd\bbm{r}_0 \, \delta\left(r_{01}-\frac{1}{2}\right)\delta\left(r_{02}-\frac{1}{2}\right),
\label{A1}
\eal
where we have taken into account that, if $r_{0i}=r_{0j}=\frac{1}{2}$, one has $r_{0k}>\frac{1}{2}$ $\forall k\neq i,j$.
Next, $r_{01}=r_{02}=\frac{1}{2}$ is compatible with $r_{12}\geq 1$ only if $r_{12}= 1$. Thus, using spherical coordinates,
\bal
Q_{N+1}^{(\frac{1}{2},2)}&=
\alpha_{\frac{1}{2}}^2\frac{d}{8}\Omega_{\frac{1}{2}}\frac{N(N-1)}{V^{N+1}}\int\dd \bbm{r}^N e^{-\beta\Phi_N(\bbm{r}^N)}\delta(r_{12}-1)\nn
&=\alpha_{\frac{1}{2}}^2\frac{d}{8}\rho^2\Omega_{\frac{1}{2}}Q_N \int\dd \bbm{r}\, \delta(r-1)g(r),
\label{A2}
\eal
where
\beq
g(r_{12})=\frac{V^{-(N-2)}}{Q_N}\int\dd\bbm{r}_3\cdots \int\dd\bbm{r}_N e^{-\beta\Phi_N(\bbm{r}^N)}
\label{A3}
\eeq
is the solvent RDF.
Finally, introducing the solvent cavity function
\beq
y(r)\equiv g(r) e^{\beta \phi(r)}
\label{yg}
\eeq
and using Eq.\ \eqref{phi12}, we obtain
\beq
Q_{N+1}^{(\frac{1}{2},2)}=\alpha_{\frac{1}{2}}^2 d^2 2^{d-3}\eta^2 y(1)Q_N\lim_{r\to 1}\left[\Theta(r-1)+\alpha \delta(r-1)\right],
\label{A4}
\eeq
where $\eta$ is defined by Eq.\ \eqref{eta}. This contribution is singular by a two-fold reason. First, the Heaviside function implies that the result depends on whether $\xi\to \frac{1}{2}$ from below or from above. Second, and more importantly, the $\delta$ function gives a divergent term. Both singularities are avoided by the choice $\alpha_{\frac{1}{2}}=0$.

\section{Virial, energy, and compressibility routes} \label{A.vce}

For systems of particles interacting through two-body central forces, the
thermodynamic functions can be evaluated in terms of the RDF $g(r)$.
In particular, the pressure $p$, the
excess internal energy per particle $u^{\rm ex}$, and the isothermal susceptibility $\chi$ are given by
\cite{H56,B74b,BH76,HM06}
\beq
 p=\rho kT-\frac{\rho^2}{2d}
   \int\dd\bbm{r} \,\frac{\partial \phi(r)}{\partial r} rg(r),\label{v.g}
\eeq
\beq
 u^{\rm ex}=\frac{\rho}{2}\int\dd\bbm{r}\, \phi(r)g(r),
 \label{e.g}
\eeq
\beq
 \chi\equiv {k_BT}\left(\frac{\partial \rho}{\partial p}\right)_T= 1+\rho\int\dd\bbm{r} \,[g(r)-1].
 \label{c.g}
 \eeq
Equations (\ref{v.g}), (\ref{e.g}), and
(\ref{c.g}) are usually known as the pressure (or virial), energy, and  compressibility equations, respectively.

For SHS fluids, the compressibility factor,
$Z\equiv p/\rho kT$, can be expressed from Eqs.\ (\ref{v.g}) and (\ref{phi22}) in terms of the
contact values of the cavity function
and its radial derivative $y'(r)$ as
\beq \label{Zv}
 Z^{(v)}(\eta,\alpha)=1+2^{d-1}\eta\{y(1)-\alpha[d y(1)+ y'(1)]\}.
\eeq
Here, the superscript $v$ specifies that the compressibility factor proceeds from
the virial equation.

In turn, the excess of internal energy per particle is related with the
compressibility factor as follows:
\beq
\rho \frac{\partial u^{\rm ex}}{\partial \rho}
 =-k_BT^2 \frac{\partial Z}{\partial T}.
\eeq
For SHS fluids, the changes of variables $\rho\rightarrow\eta$ and
$T\rightarrow\alpha$ yield
\beq \label{uZ}
\eta \frac{\partial u^{\rm ex}/\epsilon}{\partial \eta}
  =\alpha \frac{\partial Z}{\partial \alpha},
\eeq
where we have taken into account that, according to Eq.\ \eqref{alpha}, $\partial_T=-(\epsilon\alpha/k_BT^2)\partial_\alpha$.
Moreover, the excess energy can be expressed from Eq.\ (\ref{e.g}) in terms
of the cavity function using Eqs.\ (\ref{yg}) and (\ref{phi22}):
\beq \label{uy}
\frac{u^{\rm ex}}{\epsilon}=
  -d{2^{d-1}}\eta\alpha y(1).
\eeq
Integration of Eq.\ (\ref{uZ}) with (\ref{uy}) yields the compressibility factor
in the energy route as
\beq \label{Zu}
Z^{(e)}(\eta,\alpha)=Z_{\rm HS}(\eta)
          -{d2^{d-1}}\eta\int_0^\alpha\dd \alpha'\,
           \left( \frac{\partial[\eta y(1)]}{\partial\eta}\right)_{\alpha'},
\eeq
where $Z_{\rm HS}(\eta)$ is the compressibility factor for pure HS (which here remains undetermined) and in the integrand $y(1)$ is a function of $\eta$ and $\alpha'$.

As for the compressibility route, taking into account that $\chi^{-1}= \left({\partial \eta Z}/{\partial \eta}\right)_T$ and introducing the moments
\beq
H_n\equiv \int_0^\infty \dd r \,r^n h(r)
\label{Hn}
\eeq
of the total correlation function
$h(r)=g(r)-1$, one can find
\beq \label{Zc}
Z^{(c)}(\eta,\alpha) = \int_0^1\frac{\dd t}{\chi(\eta t)}
=\int_0^1 \frac {\dd t}{1+d2^d  \eta H_{d-1}(\eta t)}.
\eeq
%

\section{RDF of sticky hard rods}\label{A.shr}
The exact solution of one-dimensional ($d=1$) fluids with nearest-neighbor interactions is well known \cite{SZK53,LZ71,HC04b,S07,note_13_08}. In the case of an infinitely diluted solute particle  in a solvent, the Laplace transform
\beq
G_\xi(s)=\int_0^\infty \dd r\, e^{-rs} g_\xi(r)
\eeq
of the RDF $g_\xi(r)$  is given by
\beq
G_\xi(s) = \frac 1\rho\frac{\Psi_\xi(s+\beta p)/\Psi_\xi(\beta p)}
{1-\Psi(s+\beta p)/\Psi(\beta p)},
\label{Gxi}
\eeq
where  $\Psi(s)$ and $\Psi_\xi(s)$ are the
Laplace transforms of $e^{-\beta\phi(r)}$ (solvent-solvent interaction) and
$e^{-\beta\phi_\xi(r)}$ (solute-solvent interaction), respectively.

In the particular case of sticky hard rods, use of Eqs.\ (\ref{phi22}) and (\ref{phi12}) gives
\beq
\Psi(s)=\left(\frac 1s + \alpha \right)e^{-s},\quad
\Psi_{\xi}(s)=\left(\frac 1s + \alpha_\xi \xi\right)e^{-s\xi}.
\eeq
Moreover, the exact EOS is
\beq
\beta p=\frac{\sqrt{1+4\alpha\eta/(1-\eta)}-1}{2\alpha}.
\label{EOS_shr}
\eeq
Expansion of the right-hand side of Eq.\ \eqref{Gxi} in powers of $e^{-s}$ allows one to obtain $g_\xi(r)$ in the shells $0<r<1+\xi$, $1+\xi<r<2+\xi$, $\ldots$  In particular, if $r<1+\xi$,
\beq
g_\xi(r)=\frac{\alpha_\xi \xi\delta(r-\xi)+e^{-\beta p(r-\xi)}\Theta(r-\xi)}{\eta(\alpha_\xi\xi+1/\beta p)},\quad r<1+\xi.
\label{gxi}
\eeq
Taking into account Eqs.\ \eqref{phi12} and \eqref{yxi}, one has
\beq
y_\xi(r)=\frac{e^{-\beta p(r-\xi)}}{\eta(\alpha_\xi\xi+1/\beta p)},\quad \xi\leq r<1+\xi.
\label{gxibis}
\eeq
{}From here one easily gets Eq.\ \eqref{yyd1}.

\section{Solution of the PY equation for SHS}\label{A.shs}
In this Appendix we summarize the main results obtained from the exact solution of the PY integral equation for SHS. The reader is referred to Refs.\ \cite{B68,B74,YS93a,YS93b,PS75,B75,BT79,SYH98} for further details.

\subsection{Solvent properties}
The PY solution is expressed in terms of the Laplace transform
\beq
G(s)=\int_0^\infty \dd r\, e^{-rs} rg(r)
\eeq
of $rg(r)$. Such a solution is
\beq \label{e.Gsd3}
s^2e^sG(s)=\frac{L_0+L_1s+L_2s^2}{1-12\eta\left[
\psi_2(s)L_0+\psi_1(s)L_1+\psi_0(s)L_2 \right]},
\eeq
where
\beq \label{e.phi}
\psi_n(s)\equiv\frac{1}{s^{n+1}}\left[\sum_{m=0}^n\frac{(-s)^m}{m!}-e^{-s}\right].
\eeq
The quantities $L_0$, $L_1$, and $L_2$ are given as functions of $\eta$ and $\alpha$ by
\beq
L_0=\frac{1+2\eta}{(1-\eta)^2}-\frac{12\eta}{1-\eta}L_2,\quad
L_1=\frac{1+\eta/2}{(1-\eta)^2}-\frac{6\eta}{1-\eta}L_2,\label{L0}
\eeq
\beq \label{e.l2}
L_2=\frac{1-(1-12\alpha)\eta-K}{24\alpha(1-\eta)\eta},
\eeq
where
\beq
K\equiv\sqrt{(1-\eta)
[1-\eta(1-24\alpha+48\alpha^2)]+72\alpha^2\eta^2}.
\label{K}
\eeq

The large-$s$ behavior of $G(s)$ provides the contact values of $y(r)$ and $y'(r)$. The results are
\beq \label{yyd3}
y(1)=\frac{L_2}{\alpha},
\eeq
\bal
y'(1)=&-\frac{9\eta(1+\eta)}{2(1-\eta)^3}+\frac{12\eta(1+5\eta)}{(1-\eta)^2}L_2
-\frac{12\eta(1+11\eta)}{1-\eta}L_2^2\nn
&+144\eta^2L_2^3.
\label{yypd3}
\eal
Insertion of these expressions into Eq.\ \eqref{Zv} (with $d=3$) gives the virial equation
\bal
Z^{(v)}=&\frac{1+2\eta+3\eta^2}{(1-\eta)^2}+18\alpha\frac{\eta^2(1+\eta)}{(1-\eta)^2}\nn
&-\frac{12\eta}{1-\eta}\left[1+3\eta+4\alpha\frac{\eta(1+5\eta)}{1-\eta}\right]L_2\nn
&+48\eta^2 \left(1+\alpha\frac{1+11\eta}{1-\eta}\right)L_2^2-576\eta^3\alpha L_2^3.
\label{ZvPY}
\eal
Analogously, insertion of Eq.\ \eqref{yyd3} into Eq.\ \eqref{Zu} yields an analytical  expression for $Z^{(e)}-Z_\hs$.

The moment $H_2$ of the total correlation function [cf.\ Eq.\ \eqref{Hn}] can be obtained from the small-$s$ behavior of $G(s)$ as $s^2G(s)=1+H_1 s^2-H_2 s^3 +\mathcal{O}(s^3)$. Inserting the resulting expression of $H_2$ into $\chi=1+24\eta H_2$ [cf.\ Eq.\ \eqref{c.g}], one obtains
\beq
\frac{1}{\chi}=\frac{\left[1+2\eta-12\eta(1-\eta)L_2\right]^2}{(1-\eta)^4}.
\label{ichi}
\eeq
The compressibility factor $Z^{(c)}$ by the compressibility route is readily obtained in analytical form by application of Eq.\ \eqref{Zc}.
For conciseness, the explicit expressions of $Z^{(c)}$ and $Z^{(e)}$ will be omitted here.

As a consequence of the square root present in $K(\eta,\alpha)$ [cf.\ Eq.\ \eqref{K}], the PY solution is not physically meaningful if $\alpha>\alpha_c^{(c)}\equiv \frac{2+ \sqrt{2}}{4}\simeq 0.85355$ (or $\tau<1/12\alpha_c^{(c)}=0.09763$) and $\eta_-(\alpha)<\eta<\eta_+(\alpha)$, where
\beq
\eta_{\pm}(\alpha)=\frac{1-12\alpha+24\alpha^2 \pm 6\alpha
 \sqrt{2-16\alpha(1-\alpha)}}{1-24\alpha+120\alpha^2}.
\eeq
In the limit $\alpha\to\alpha_c^{(c)}$ one has $\eta_{\pm}\to\eta_c^{(c)}=(3 \sqrt{2}-4)/2\simeq 0.121\,32$.
It can be easily checked that the right-hand side of Eq.\ \eqref{ichi} vanishes at $(\eta,\alpha)=(\eta_c^{(c)},\alpha_c^{(c)})$. This implies that $(\eta_c^{(c)},\alpha_c^{(c)})$ is the critical point in the compressibility route.

{As an extra route, Barboy and Tenne \cite{BT79} applied the so-called ZS theorem \cite{BT76} to the PY solution for SHS fluids. According to this ZS route,  the excess chemical potential is expressed as}
\beq
{\beta \mu^\text{ex}=\ln y_\text{reg},}
\label{ZS}
\eeq
{where}
\beq
{y_\text{reg}=\frac{\left[1-4\eta-(1-\eta-K)/2\alpha\right]^2}{(1-\eta)^4}}
\label{yreg}
\eeq
{is the regular part of the cavity function at $r=0$. The associated compressibility factor is then obtained from  the thermodynamic relation \eqref{Zmu}, i.e.,}
\beq
{Z^{(\text{ZS})}(\eta,\alpha)=1+\ln y_\text{reg}(\eta,\alpha)-\int_0^1 dt\, \ln y_\text{reg}(\eta t,\alpha)}.
\label{ZZS}
\eeq

\subsection{Solute-solvent RDF}
{}From the exact solution of the PY equation for an SHS binary mixture \cite{PS75,SYH98} one can take the limit where one of the species (the solute) is infinitely dilute. As a result, the Laplace transform
\beq
G_\xi(s)=\int_0^\infty \dd r\, e^{-rs} rg_\xi(r)
\eeq
of $rg_\xi(r)$ is given by
\beq
\label{e.Gxi}
s^2e^{s\xi}G_\xi(s)=\frac{L_0+L_1^{(\xi)}s+L_2^{(\xi)}s^2}{1-12\eta\left[
\psi_2(s)L_0+\psi_1(s)L_1+\psi_0(s)L_2 \right]},
\eeq
where
\beq
L_1^{(\xi)}=\frac{\xi+\eta(2\xi-3/2)}{(1-\eta)^2}
    -\frac{6\eta(2\xi-1)}{1-\eta}L_2, \label{P1}
    \eeq
    \beq
    L_2^{(\xi)}=\left(\frac{1}{\alpha_\xi\xi}+\frac{6\eta}{1-\eta}
    -12\eta L_2 \right)^{-1} L_1^{(\xi)}.\label{P2}
\eeq
From the large-$s$ behavior of $G_\xi(s)$ we can obtain  the contact values of $y_\xi(r)$ and $y_\xi'(r)$ as
\beq \label{yyd3x}
y_\xi(\xi) = \frac{L_2^{(\xi)}}{\alpha_\xi\xi^2},
\eeq
\bal
\label{yyd3xc}
\xi y'_\xi(\xi) =& 12\eta L_2^{(\xi)}  \left[3\eta\left(L_0 -2L_1 +2L_2\right)^2-L_0+L_1\right]
\nn
  & +  L_0 +6\eta L_1^{(\xi)} (L_0 -2L_1 +2L_2)-y_\xi(\xi).
\eal

\bibliographystyle{apsrev}

\bibliography{D:/Dropbox/Public/bib_files/liquid}

\begin{thebibliography}{41}
\expandafter\ifx\csname natexlab\endcsname\relax\def\natexlab#1{#1}\fi
\expandafter\ifx\csname bibnamefont\endcsname\relax
  \def\bibnamefont#1{#1}\fi
\expandafter\ifx\csname bibfnamefont\endcsname\relax
  \def\bibfnamefont#1{#1}\fi
\expandafter\ifx\csname citenamefont\endcsname\relax
  \def\citenamefont#1{#1}\fi
\expandafter\ifx\csname url\endcsname\relax
  \def\url#1{\texttt{#1}}\fi
\expandafter\ifx\csname urlprefix\endcsname\relax\def\urlprefix{URL }\fi
\providecommand{\bibinfo}[2]{#2}
\providecommand{\eprint}[2][]{\url{#2}}

\bibitem[{\citenamefont{Onsager}(1933)}]{O33}
\bibinfo{author}{\bibfnamefont{L.}~\bibnamefont{Onsager}},
  \bibinfo{journal}{Chem. Rev.} \textbf{\bibinfo{volume}{13}},
  \bibinfo{pages}{73} (\bibinfo{year}{1933}).

\bibitem[{\citenamefont{Kirkwood}(1935)}]{K35}
\bibinfo{author}{\bibfnamefont{J.~G.} \bibnamefont{Kirkwood}},
  \bibinfo{journal}{J. Chem. Phys.} \textbf{\bibinfo{volume}{3}},
  \bibinfo{pages}{300} (\bibinfo{year}{1935}).

\bibitem[{\citenamefont{Hill}(1956)}]{H56}
\bibinfo{author}{\bibfnamefont{T.~L.} \bibnamefont{Hill}},
  \emph{\bibinfo{title}{Statistical Mechanics}}
  (\bibinfo{publisher}{McGraw-Hill}, \bibinfo{address}{New York},
  \bibinfo{year}{1956}).

\bibitem[{\citenamefont{Reichl}(1980)}]{R80}
\bibinfo{author}{\bibfnamefont{L.~E.} \bibnamefont{Reichl}},
  \emph{\bibinfo{title}{A Modern Course in Statistical Physics}}
  (\bibinfo{publisher}{University of Texas Press}, \bibinfo{address}{Austin},
  \bibinfo{year}{1980}), \bibinfo{edition}{1st} ed.

\bibitem[{\citenamefont{Hansen and McDonald}(2006)}]{HM06}
\bibinfo{author}{\bibfnamefont{J.-P.} \bibnamefont{Hansen}} \bibnamefont{and}
  \bibinfo{author}{\bibfnamefont{I.~R.} \bibnamefont{McDonald}},
  \emph{\bibinfo{title}{{Theory of Simple Liquids}}}
  (\bibinfo{publisher}{Academic}, \bibinfo{address}{London},
  \bibinfo{year}{2006}).

\bibitem[{\citenamefont{Salsburg et~al.}(1953)\citenamefont{Salsburg, Zwanzig,
  and Kirkwood}}]{SZK53}
\bibinfo{author}{\bibfnamefont{Z.~W.} \bibnamefont{Salsburg}},
  \bibinfo{author}{\bibfnamefont{R.~W.} \bibnamefont{Zwanzig}},
  \bibnamefont{and} \bibinfo{author}{\bibfnamefont{J.~G.}
  \bibnamefont{Kirkwood}}, \bibinfo{journal}{J. Chem. Phys.}
  \textbf{\bibinfo{volume}{21}}, \bibinfo{pages}{1098} (\bibinfo{year}{1953}).

\bibitem[{\citenamefont{Reiss et~al.}(1959)\citenamefont{Reiss, Frisch, and
  Lebowitz}}]{RFL59}
\bibinfo{author}{\bibfnamefont{H.}~\bibnamefont{Reiss}},
  \bibinfo{author}{\bibfnamefont{H.~L.} \bibnamefont{Frisch}},
  \bibnamefont{and} \bibinfo{author}{\bibfnamefont{J.~L.}
  \bibnamefont{Lebowitz}}, \bibinfo{journal}{J. Chem. Phys.}
  \textbf{\bibinfo{volume}{31}}, \bibinfo{pages}{369} (\bibinfo{year}{1959}).

\bibitem[{\citenamefont{Lebowitz et~al.}(1965)\citenamefont{Lebowitz, Helfand,
  and Praestgaard}}]{LHP65}
\bibinfo{author}{\bibfnamefont{J.~L.} \bibnamefont{Lebowitz}},
  \bibinfo{author}{\bibfnamefont{E.}~\bibnamefont{Helfand}}, \bibnamefont{and}
  \bibinfo{author}{\bibfnamefont{E.}~\bibnamefont{Praestgaard}},
  \bibinfo{journal}{J. Chem. Phys.} \textbf{\bibinfo{volume}{43}},
  \bibinfo{pages}{774} (\bibinfo{year}{1965}).

\bibitem[{\citenamefont{Mandell and Reiss}(1975)}]{MR75}
\bibinfo{author}{\bibfnamefont{M.}~\bibnamefont{Mandell}} \bibnamefont{and}
  \bibinfo{author}{\bibfnamefont{H.}~\bibnamefont{Reiss}}, \bibinfo{journal}{J.
  Stat. Phys.} \textbf{\bibinfo{volume}{13}}, \bibinfo{pages}{113}
  (\bibinfo{year}{1975}).

\bibitem[{\citenamefont{Heying and Corti}(2004)}]{HC04b}
\bibinfo{author}{\bibfnamefont{M.}~\bibnamefont{Heying}} \bibnamefont{and}
  \bibinfo{author}{\bibfnamefont{D.}~\bibnamefont{Corti}}, \bibinfo{journal}{J.
  Phys. Chem. B} \textbf{\bibinfo{volume}{108}}, \bibinfo{pages}{19756}
  (\bibinfo{year}{2004}).

\bibitem[{\citenamefont{Stillinger et~al.}(2006)\citenamefont{Stillinger,
  Debenedetti, and Chatterjee}}]{SDC06}
\bibinfo{author}{\bibfnamefont{F.~H.} \bibnamefont{Stillinger}},
  \bibinfo{author}{\bibfnamefont{P.~G.} \bibnamefont{Debenedetti}},
  \bibnamefont{and}
  \bibinfo{author}{\bibfnamefont{S.}~\bibnamefont{Chatterjee}},
  \bibinfo{journal}{J. Chem. Phys.} \textbf{\bibinfo{volume}{125}},
  \bibinfo{pages}{204504} (\bibinfo{year}{2006}).

\bibitem[{\citenamefont{Santos}(2012)}]{S12b}
\bibinfo{author}{\bibfnamefont{A.}~\bibnamefont{Santos}},
  \bibinfo{journal}{Phys. Rev. Lett.} \textbf{\bibinfo{volume}{109}},
  \bibinfo{pages}{120601} (\bibinfo{year}{2012}).

\bibitem[{\citenamefont{Santos and Rohrmann}(2013)}]{SR13}
\bibinfo{author}{\bibfnamefont{A.}~\bibnamefont{Santos}} \bibnamefont{and}
  \bibinfo{author}{\bibfnamefont{R.~D.} \bibnamefont{Rohrmann}},
  \bibinfo{journal}{Phys. Rev. E} \textbf{\bibinfo{volume}{87}},
  \bibinfo{pages}{052138} (\bibinfo{year}{2013}).

\bibitem[{\citenamefont{Beltr\'an-Heredia and Santos}(2014)}]{BS14}
\bibinfo{author}{\bibfnamefont{E.}~\bibnamefont{Beltr\'an-Heredia}}
  \bibnamefont{and} \bibinfo{author}{\bibfnamefont{A.}~\bibnamefont{Santos}},
  \bibinfo{journal}{J. Chem. Phys.} \textbf{\bibinfo{volume}{140}},
  \bibinfo{pages}{134507} (\bibinfo{year}{2014}).

\bibitem[{\citenamefont{Baxter}(1968)}]{B68}
\bibinfo{author}{\bibfnamefont{R.~J.} \bibnamefont{Baxter}},
  \bibinfo{journal}{J. Chem. Phys.} \textbf{\bibinfo{volume}{49}},
  \bibinfo{pages}{2770} (\bibinfo{year}{1968}).

\bibitem[{\citenamefont{Barboy}(1974)}]{B74}
\bibinfo{author}{\bibfnamefont{B.}~\bibnamefont{Barboy}}, \bibinfo{journal}{J.
  Chem. Phys.} \textbf{\bibinfo{volume}{61}}, \bibinfo{pages}{3194}
  (\bibinfo{year}{1974}).

\bibitem[{\citenamefont{Chen et~al.}(1994)\citenamefont{Chen, Rouch, Sciortino,
  and Tartaglia}}]{CRST94}
\bibinfo{author}{\bibfnamefont{S.~H.} \bibnamefont{Chen}},
  \bibinfo{author}{\bibfnamefont{J.}~\bibnamefont{Rouch}},
  \bibinfo{author}{\bibfnamefont{F.}~\bibnamefont{Sciortino}},
  \bibnamefont{and}
  \bibinfo{author}{\bibfnamefont{P.}~\bibnamefont{Tartaglia}},
  \bibinfo{journal}{J. Phys.: Condens. Matter} \textbf{\bibinfo{volume}{6}},
  \bibinfo{pages}{109855} (\bibinfo{year}{1994}).

\bibitem[{\citenamefont{Verduin and Dhont}(1995)}]{VD95}
\bibinfo{author}{\bibfnamefont{H.}~\bibnamefont{Verduin}} \bibnamefont{and}
  \bibinfo{author}{\bibfnamefont{J.~K.~G.} \bibnamefont{Dhont}},
  \bibinfo{journal}{J. Colloid Interf. Sci.} \textbf{\bibinfo{volume}{172}},
  \bibinfo{pages}{425} (\bibinfo{year}{1995}).

\bibitem[{\citenamefont{Rosenbaum et~al.}(1996)\citenamefont{Rosenbaum, Zamora,
  and Zukoski}}]{RZZ96}
\bibinfo{author}{\bibfnamefont{D.}~\bibnamefont{Rosenbaum}},
  \bibinfo{author}{\bibfnamefont{P.~C.} \bibnamefont{Zamora}},
  \bibnamefont{and} \bibinfo{author}{\bibfnamefont{C.~F.}
  \bibnamefont{Zukoski}}, \bibinfo{journal}{Phys. Rev. Lett.}
  \textbf{\bibinfo{volume}{76}}, \bibinfo{pages}{150} (\bibinfo{year}{1996}).

\bibitem[{\citenamefont{Pontoni et~al.}(2003)\citenamefont{Pontoni, Finet,
  Narayanan, and Rennie}}]{PFNR03}
\bibinfo{author}{\bibfnamefont{D.}~\bibnamefont{Pontoni}},
  \bibinfo{author}{\bibfnamefont{S.}~\bibnamefont{Finet}},
  \bibinfo{author}{\bibfnamefont{T.}~\bibnamefont{Narayanan}},
  \bibnamefont{and} \bibinfo{author}{\bibfnamefont{A.~R.}
  \bibnamefont{Rennie}}, \bibinfo{journal}{J. Chem. Phys.}
  \textbf{\bibinfo{volume}{119}}, \bibinfo{pages}{6157} (\bibinfo{year}{2003}).

\bibitem[{\citenamefont{Buzzaccaro et~al.}(2007)\citenamefont{Buzzaccaro,
  Rusconi, and Piazza}}]{BRP07}
\bibinfo{author}{\bibfnamefont{S.}~\bibnamefont{Buzzaccaro}},
  \bibinfo{author}{\bibfnamefont{R.}~\bibnamefont{Rusconi}}, \bibnamefont{and}
  \bibinfo{author}{\bibfnamefont{R.}~\bibnamefont{Piazza}},
  \bibinfo{journal}{Phys. Rev. Lett.} \textbf{\bibinfo{volume}{99}},
  \bibinfo{pages}{098301} (\bibinfo{year}{2007}).

\bibitem[{\citenamefont{Piazza et~al.}(1998)\citenamefont{Piazza, Peyre, and
  Degiorgio}}]{PPD98}
\bibinfo{author}{\bibfnamefont{R.}~\bibnamefont{Piazza}},
  \bibinfo{author}{\bibfnamefont{V.}~\bibnamefont{Peyre}}, \bibnamefont{and}
  \bibinfo{author}{\bibfnamefont{V.}~\bibnamefont{Degiorgio}},
  \bibinfo{journal}{Phys. Rev. E} \textbf{\bibinfo{volume}{58}},
  \bibinfo{pages}{R2733} (\bibinfo{year}{1998}).

\bibitem[{\citenamefont{Noro and Frenkel}(2000)}]{NF00}
\bibinfo{author}{\bibfnamefont{M.~G.} \bibnamefont{Noro}} \bibnamefont{and}
  \bibinfo{author}{\bibfnamefont{D.}~\bibnamefont{Frenkel}},
  \bibinfo{journal}{J. Chem. Phys.} \textbf{\bibinfo{volume}{113}},
  \bibinfo{pages}{2941} (\bibinfo{year}{2000}).

\bibitem[{\citenamefont{Maestre et~al.}(2013)\citenamefont{Maestre, Fantoni,
  Giacometti, and Santos}}]{MFGS13}
\bibinfo{author}{\bibfnamefont{M.~A.~G.} \bibnamefont{Maestre}},
  \bibinfo{author}{\bibfnamefont{R.}~\bibnamefont{Fantoni}},
  \bibinfo{author}{\bibfnamefont{A.}~\bibnamefont{Giacometti}},
  \bibnamefont{and} \bibinfo{author}{\bibfnamefont{A.}~\bibnamefont{Santos}},
  \bibinfo{journal}{J. Chem. Phys.} \textbf{\bibinfo{volume}{138}},
  \bibinfo{pages}{094904} (\bibinfo{year}{2013}).

\bibitem[{\citenamefont{Perram and Smith}(1975)}]{PS75}
\bibinfo{author}{\bibfnamefont{J.~W.} \bibnamefont{Perram}} \bibnamefont{and}
  \bibinfo{author}{\bibfnamefont{E.~R.} \bibnamefont{Smith}},
  \bibinfo{journal}{Chem. Phys. Lett.} \textbf{\bibinfo{volume}{35}},
  \bibinfo{pages}{138} (\bibinfo{year}{1975}).

\bibitem[{\citenamefont{Barboy and Tenne}(1979)}]{BT79}
\bibinfo{author}{\bibfnamefont{B.}~\bibnamefont{Barboy}} \bibnamefont{and}
  \bibinfo{author}{\bibfnamefont{R.}~\bibnamefont{Tenne}},
  \bibinfo{journal}{Chem. Phys.} \textbf{\bibinfo{volume}{38}},
  \bibinfo{pages}{369} (\bibinfo{year}{1979}).

\bibitem[{\citenamefont{Barboy and Tenne}(1976)}]{BT76}
\bibinfo{author}{\bibfnamefont{B.}~\bibnamefont{Barboy}} \bibnamefont{and}
  \bibinfo{author}{\bibfnamefont{R.}~\bibnamefont{Tenne}},
  \bibinfo{journal}{Mol. Phys.} \textbf{\bibinfo{volume}{31}},
  \bibinfo{pages}{1749} (\bibinfo{year}{1976}).

\bibitem[{\citenamefont{Miller and Frenkel}(2004)}]{MF04}
\bibinfo{author}{\bibfnamefont{M.~A.} \bibnamefont{Miller}} \bibnamefont{and}
  \bibinfo{author}{\bibfnamefont{D.}~\bibnamefont{Frenkel}},
  \bibinfo{journal}{J. Chem. Phys.} \textbf{\bibinfo{volume}{121}},
  \bibinfo{pages}{535} (\bibinfo{year}{2004}).

\bibitem[{\citenamefont{Post and Glandt}(1986)}]{PG86}
\bibinfo{author}{\bibfnamefont{A.~J.} \bibnamefont{Post}} \bibnamefont{and}
  \bibinfo{author}{\bibfnamefont{E.~D.} \bibnamefont{Glandt}},
  \bibinfo{journal}{J. Chem. Phys.} \textbf{\bibinfo{volume}{84}},
  \bibinfo{pages}{4585} (\bibinfo{year}{1986}).

\bibitem[{\citenamefont{Santos}(2005)}]{S05}
\bibinfo{author}{\bibfnamefont{A.}~\bibnamefont{Santos}}, \bibinfo{journal}{J.
  Chem. Phys.} \textbf{\bibinfo{volume}{123}}, \bibinfo{pages}{104102}
  (\bibinfo{year}{2005}).

\bibitem[{\citenamefont{Santos}(2006)}]{S06}
\bibinfo{author}{\bibfnamefont{A.}~\bibnamefont{Santos}},
  \bibinfo{journal}{Mol. Phys.} \textbf{\bibinfo{volume}{104}},
  \bibinfo{pages}{3411} (\bibinfo{year}{2006}).

\bibitem[{\citenamefont{Watts et~al.}(1971)\citenamefont{Watts, Henderson, and
  Baxter}}]{WHB71}
\bibinfo{author}{\bibfnamefont{R.~O.} \bibnamefont{Watts}},
  \bibinfo{author}{\bibfnamefont{D.}~\bibnamefont{Henderson}},
  \bibnamefont{and} \bibinfo{author}{\bibfnamefont{R.~J.}
  \bibnamefont{Baxter}}, \bibinfo{journal}{Adv. Chem. Phys.}
  \textbf{\bibinfo{volume}{21}}, \bibinfo{pages}{421} (\bibinfo{year}{1971}).

\bibitem[{\citenamefont{Balescu}(1974)}]{B74b}
\bibinfo{author}{\bibfnamefont{R.}~\bibnamefont{Balescu}},
  \emph{\bibinfo{title}{Equilibrium and Nonequilibrium Statistical Mechanics}}
  (\bibinfo{publisher}{Wiley}, \bibinfo{address}{New York},
  \bibinfo{year}{1974}).

\bibitem[{\citenamefont{Barker and Henderson}(1976)}]{BH76}
\bibinfo{author}{\bibfnamefont{J.~A.} \bibnamefont{Barker}} \bibnamefont{and}
  \bibinfo{author}{\bibfnamefont{D.}~\bibnamefont{Henderson}},
  \bibinfo{journal}{Rev. Mod. Phys.} \textbf{\bibinfo{volume}{48}},
  \bibinfo{pages}{587} (\bibinfo{year}{1976}).

\bibitem[{\citenamefont{Lebowitz and Zomick}(1971)}]{LZ71}
\bibinfo{author}{\bibfnamefont{J.~L.} \bibnamefont{Lebowitz}} \bibnamefont{and}
  \bibinfo{author}{\bibfnamefont{D.}~\bibnamefont{Zomick}},
  \bibinfo{journal}{J. Chem. Phys.} \textbf{\bibinfo{volume}{54}},
  \bibinfo{pages}{3335} (\bibinfo{year}{1971}).

\bibitem[{\citenamefont{Santos}(2007)}]{S07}
\bibinfo{author}{\bibfnamefont{A.}~\bibnamefont{Santos}},
  \bibinfo{journal}{Phys. Rev. E} \textbf{\bibinfo{volume}{76}},
  \bibinfo{pages}{062201} (\bibinfo{year}{2007}).

\bibitem[{not()}]{note_13_08}
\bibinfo{note}{See also A. Santos, ``Radial Distribution Function for Sticky
  Hard Rods'',
  \url{http://demonstrations.wolfram.com/RadialDistributionFunctionForStickyHardRods/},
  Wolfram Demonstrations Project}.

\bibitem[{\citenamefont{Yuste and Santos}(1993{\natexlab{a}})}]{YS93a}
\bibinfo{author}{\bibfnamefont{S.~B.} \bibnamefont{Yuste}} \bibnamefont{and}
  \bibinfo{author}{\bibfnamefont{A.}~\bibnamefont{Santos}},
  \bibinfo{journal}{J. Stat. Phys.} \textbf{\bibinfo{volume}{72}},
  \bibinfo{pages}{703} (\bibinfo{year}{1993}{\natexlab{a}}).

\bibitem[{\citenamefont{Yuste and Santos}(1993{\natexlab{b}})}]{YS93b}
\bibinfo{author}{\bibfnamefont{S.~B.} \bibnamefont{Yuste}} \bibnamefont{and}
  \bibinfo{author}{\bibfnamefont{A.}~\bibnamefont{Santos}},
  \bibinfo{journal}{Phys. Rev. E} \textbf{\bibinfo{volume}{48}},
  \bibinfo{pages}{4599} (\bibinfo{year}{1993}{\natexlab{b}}).

\bibitem[{\citenamefont{Barboy}(1975)}]{B75}
\bibinfo{author}{\bibfnamefont{B.}~\bibnamefont{Barboy}},
  \bibinfo{journal}{Chem. Phys.} \textbf{\bibinfo{volume}{11}},
  \bibinfo{pages}{357} (\bibinfo{year}{1975}).

\bibitem[{\citenamefont{Santos et~al.}(1998)\citenamefont{Santos, Yuste, and
  {L\'opez de Haro}}}]{SYH98}
\bibinfo{author}{\bibfnamefont{A.}~\bibnamefont{Santos}},
  \bibinfo{author}{\bibfnamefont{S.~B.} \bibnamefont{Yuste}}, \bibnamefont{and}
  \bibinfo{author}{\bibfnamefont{M.}~\bibnamefont{{L\'opez de Haro}}},
  \bibinfo{journal}{J. Chem. Phys.} \textbf{\bibinfo{volume}{109}},
  \bibinfo{pages}{6814} (\bibinfo{year}{1998}).

\end{thebibliography}

\end{document}